\documentclass[11pt]{article}

\usepackage[a4paper,margin=1in]{geometry}
\usepackage{amsmath,amssymb,bm}
\usepackage{graphicx}
\usepackage{booktabs}
\usepackage{array}
\usepackage{multirow}
\usepackage{float}
\usepackage{xurl}
\usepackage[numbers,sort&compress]{natbib}
\usepackage[hidelinks]{hyperref}
\usepackage{microtype}
\usepackage{setspace}
\newenvironment{ruledtabular}{\centering}{}

\begin{document}

\title{Direct Cultivation of Entangled $|CS\rangle$ Magic States}
\author{
Gunsik Min$^{1,*}$ and Jun Heo$^{1}$\\[0.5em]
\small $^{1}$School of Electrical Engineering, Korea University,
Seoul 02841, Republic of Korea\\[0.35em]
\small $^{*}$Correspondence: mgs3351@korea.ac.kr
}
\date{}
\maketitle

\begin{abstract}
Magic-state cultivation has largely targeted single-qubit non-Clifford
resources. We show that the entangled state $|CS\rangle=CS|++\rangle$ can be
cultivated directly rather than synthesized from three $T$-type resources.
Two commuting Hermitian Clifford involutions project onto four Pauli-frame
branches, so no ideal branch is discarded, while a minimal six-bit $[6,2,4]$
record protects their history. The small-code state cannot itself be certified:
late two-fault patterns evade its record, placing the certified boundary after
surface-code growth. Under a fixed decoder that sees 136 operational detector
bits per patch and withholds 168, exact audits find no accepted closed-boundary
logical failure through fault order two, while order-three mechanisms exist.
Escaping from the small code directly into a distance-3 patch and growing it to
distance 13 in a single round keeps this certificate. Compared with a re-decoded
three-$T$ comparator matched to the same binary logical-error target, the
direct route uses $2\%$ fewer active locations at $p=5\times10^{-4}$ and
$16$--$28\%$ more at $8\times10^{-4}$--$10^{-3}$, without charging the
comparator for $T$-state storage. Counted in qubit-ticks under one rule for
both routes, it uses $21$--$39\%$ less spacetime. Adding idle-memory noise to
both routes under one rule and one optimized schedule raises the direct route's
cost more, to $8\%$ and $46\%$ above the comparator at $5\times10^{-4}$ and
$10^{-3}$, because that cost is limited by acceptance. The result is architectural: an
entangled non-Clifford state can itself be cultivated with a certified output
interface.
\end{abstract}

\section{Introduction}
\label{sec:introduction}

Fault-tolerant quantum computing commonly relies on non-stabilizer resource states to supply logical non-Clifford operations~\cite{BravyiKitaev2005,EastinKnill2009}. In surface-code architectures~\cite{Fowler2012}, preparing those resources can constitute a substantial fraction of the computational overhead, motivating extensive work on magic-state distillation and factory design~\cite{FowlerDevittJones2013,OGormanCampbell2017,Litinski2019MSD,GidneyEkera2021}. Most such pipelines use the single-qubit $|T\rangle$ state as a standard intermediate resource: one first prepares several $T$-type resources and then synthesizes the desired multiqubit non-Clifford operation. The natural factory output, however, need not be single-qubit. Direct preparation or joint distillation of multiqubit resources can merge part of this distill-then-synthesize pipeline~\cite{Eastin2013Toffoli,Jones2013Toffoli,PaetznickReichardt2013,CampbellHoward2017,CampbellHoward2018}; dedicated $|CCZ\rangle$ factories provide a prominent example~\cite{GidneyFowler2019}. Comparative studies of distillation and code switching further show that the preferred route to universality depends on the architecture and resource metric~\cite{BeverlandKubicaSvore2021,Bombin2015,DaguerreKim2025}.

Magic-state cultivation offers a complementary approach in which a non-Clifford state is progressively verified while its code protection is increased. The original proposal targeted $|T\rangle$ states using non-Pauli logical checks, postselection, and escape into a larger surface-code patch~\cite{GidneyShuttyJones2024}. Recent work has explored alternative cultivation codes and growth interfaces~\cite{Chen2026}, direct surface-code cultivation with flexible connectivity~\cite{Vaknin2026}, fold-transversal surface-code constructions~\cite{Sahay2026}, surface-code cultivation with only two-qubit gates~\cite{Claes2025}, and cultivation of higher-level single-qubit states~\cite{ChenFonsecaSornborger2026}, together with renewed work on low-overhead code switching and efficient simulation of logical magic-state preparation~\cite{DaguerreKim2025,SurtiDaguerreKim2026}. Cultivation has also been demonstrated on a superconducting processor~\cite{Rosenfeld2026}. Entangled-resource cultivation has been considered for logical $|CX\rangle$ states~\cite{Vaknin2026} and, concurrently with this work, for logical $|CCZ\rangle$ states on bivariate bicycle and tricycle codes with native non-Clifford gates~\cite{HirasakiLee2026}. The question addressed here is therefore more specific: can the resource $|CS\rangle$ be cultivated directly using joint logical checks that resolve its branch label, with the branch history and code-growth interface protected as part of the same factory?

\begin{figure}[H]
    \centering
    \includegraphics[width=0.92\textwidth]{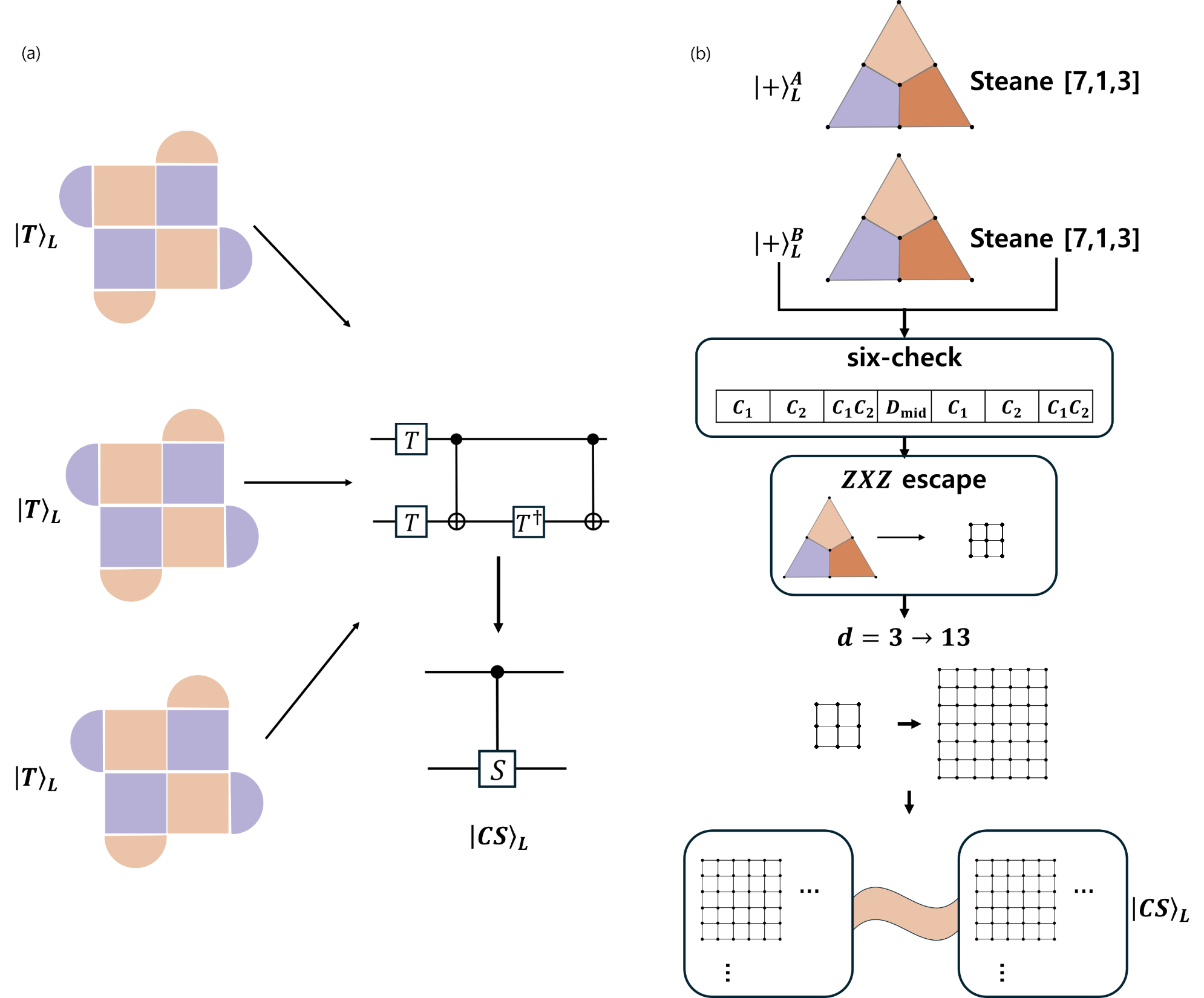}
    \caption{\textbf{Architecture overview and indirect comparator.}
    (a) The indirect route prepares three logical $|T\rangle_L$ resources and converts them to $|CS\rangle_L$ with a Clifford+$T$ circuit.
    (b) The direct route starts from two Steane $[[7,1,3]]$ blocks in $|+\rangle_L$, applies the six-check sequence $C_1,C_2,C_{12}\,|\,D_{\mathrm{mid}}\,|\,C_1,C_2,C_{12}$, and immediately transfers the accepted state through the Steane-to-$d=3$ escape and direct $d=3\!\rightarrow\!13$ surface-code growth. The three-$T$ route is used only as a resource comparator.}
    \label{fig:protocol_overview}
\end{figure}

We address this question for the controlled-$S$ resource
\begin{equation}
    |CS\rangle \equiv CS|++\rangle
    = \frac{|00\rangle+|01\rangle+|10\rangle+i|11\rangle}{2}.
    \label{eq:cs_state_intro}
\end{equation}
An indirect implementation can synthesize the controlled-$S$ operation from three $T$-type resources, for example through
\begin{equation}
    CS=(T_A\otimes T_B)\,\mathrm{CNOT}_{A\rightarrow B}
    (I\otimes T_B^{\dagger})\,\mathrm{CNOT}_{A\rightarrow B}.
    \label{eq:cs_3t_identity_intro}
\end{equation}
The $|CS\rangle$ state therefore provides a compact test of whether an entangled resource can be produced by one cultivation process rather than by preparing three independent $T$-type resources and synthesizing them afterward. Figure~\ref{fig:protocol_overview} summarizes the direct architecture and the indirect three-$T$ route used later as a resource comparator.

The construction makes four contributions. First, two commuting Hermitian Clifford involutions have rank-one joint eigenspaces and project onto four branches that differ only by Pauli-frame updates, so no ideal branch is discarded. Second, the branch history is protected by a six-check $[6,2,4]$ classical record. Six bits are the minimum length for four binary branch labels at pairwise Hamming distance four, and this suppresses valid-to-valid record aliasing relative to the five-check $[5,2,3]$ baseline. Third, the small-code cultivation core is \emph{not} a composable certified output: an explicit late two-fault pattern can share a syndrome with a single-qubit error and become logical after recovery. Code growth must therefore be included inside the certified gadget. Fourth, the resulting downstream interface separates 136 operational detector bits per patch from 168 future-validation bits that are withheld from the decoder. The broader correspondence between Clifford-parity measurement schedules and binary record codes is developed separately~\cite{MinCodedClifford2026}; the complete $|CS\rangle$ construction, composability analysis, decoder, and resource evaluation are established here.

We realize the protocol on two Steane $[[7,1,3]]$ blocks with verified CAT readout, midpoint Steane error detection, controlled logical checks, immediate escape to a distance-3 surface-code patch, and direct growth to distance 13. The practical decoder resolves certified low-order cases exactly, then uses a precomputed higher-order catalogue and four-coset BP+OSD for unresolved records. Under the fixed decoder and stated active-location stochastic-Pauli model, exact enumeration finds no accepted closed-boundary logical-failure mechanism through total fault order two, while explicit order-three mechanisms exist. We therefore describe the accepted closed-boundary channel as having fault order three rather than claiming an unconditional delivered-state distance theorem.

The factory is evaluated from $p=10^{-4}$ to $10^{-3}$. The escape-and-growth interface is the decisive cost term, and compressing it to a distance-3 escape followed by one $d=3\rightarrow13$ growth round lowers the accepted-output cost by 49--53\% relative to the earlier staged Direct schedule and by 21--25\% relative to a distance-5 interface. It also roughly halves the leading logical-error coefficient relative to that distance-5 interface, estimated with the same method. We compare the factory with three cultivated $T$ states using a common primitive convention, noise model, early-abort rule, and output boundary, after re-decoding the public three-$T$ circuit. Matched to the Direct binary-LER target and with the decoder's abstention threshold selected on independent calibration data, Direct CS uses 2\% fewer active locations than the comparator at $p=5\times10^{-4}$. At $8\times10^{-4}$--$10^{-3}$ it uses 16--28\% more, and it becomes the cheaper route once holding $T$ states for assembly costs more than 0.9--1.4 distance-13 patch-rounds per output. Counted in qubit-ticks under one rule for both routes, Direct CS uses 21--39\% less spacetime. Adding idle-memory noise to both routes under a common rule and an idle-minimizing schedule, Direct CS uses 8\% and 46\% more active locations than the comparator at $5\times10^{-4}$ and $10^{-3}$, and it remains cheaper once $T$ states must be held for more than 0.4 and 2.6 patch-rounds. The resulting contribution is therefore architectural: it demonstrates that an entangled non-Clifford resource can itself be cultivated with a protected branch record and an explicitly certified output interface.

\paragraph*{Main results and assumptions.}
\begin{itemize}
\item \emph{Construction.} Two Steane blocks, six joint logical checks with a $[6,2,4]$-protected branch record, a distance-3 surface-code escape, and one faultable growth round to distance 13 (3,762 potential active locations).
\item \emph{Certificate.} Under the stated active-location stochastic-Pauli model and the fixed decoder, which sees 136 operational detector bits per patch and withholds 168, no accepted closed-boundary logical failure occurs through total fault order two. Order-three failures occur, with $A_3^{\rm bin}=9.2(1.6)\times10^5$ at the default threshold.
\item \emph{Cost.} The cost per accepted output is 5,056--6,341 active locations at $p=5\times10^{-4}$--$10^{-3}$ at the calibrated threshold. Relative to a re-decoded three-$T$ comparator under the same conventions, the cost ratio is 0.98--1.28 without $T$-state storage, and Direct CS is cheaper once the comparator must hold $T$ states for more than 1.4 distance-13 patch-rounds.
\item \emph{Idle noise.} With idle-memory noise added to both routes under one rule and an idle-minimizing schedule, the ratio becomes 1.08 and 1.46.
\item \emph{Not modelled.} Leakage, coherent and correlated faults, operation durations, routing, and the logical consumer of the resource.
\end{itemize}

\section{Results}

\subsection{Direct $|CS\rangle$ Cultivation}
\label{sec:protocol}

\subsubsection{Logical Projection}
\label{subsec:logical_projection}

The cultivation target is the two-qubit non-Clifford state
\begin{equation}
    |CS\rangle = CS|++\rangle
    = \frac{|00\rangle+|01\rangle+|10\rangle+i|11\rangle}{2}.
    \label{eq:cs_target}
\end{equation}
Our construction is organized around the logical operators
\begin{align}
    C_1 &= X_A S_B CZ_{AB},
    \label{eq:C1}\\
    C_2 &= X_B S_A CZ_{AB}.
    \label{eq:C2}
\end{align}
Although these operators contain the phase gate $S$, the complete products in Eqs.~\eqref{eq:C1}--\eqref{eq:C2} are Hermitian Clifford involutions. Direct evaluation gives
\begin{equation}
    C_1^\dagger=C_1,\quad C_2^\dagger=C_2,\quad
    C_1^2=C_2^2=I,\quad [C_1,C_2]=0,
    \label{eq:check_properties}
\end{equation}
and the target state is stabilized by both checks,
\begin{equation}
    C_1|CS\rangle=|CS\rangle,\qquad
    C_2|CS\rangle=|CS\rangle.
    \label{eq:CS_stabilized}
\end{equation}
Thus $|CS\rangle$ can be prepared by a joint logical projection rather than by first preparing three independent $T$-type resources.

Let $r_1,r_2\in\{+1,-1\}$ denote the outcomes of the two logical measurements and define
\begin{equation}
    a=\frac{1-r_1}{2},\qquad b=\frac{1-r_2}{2}.
    \label{eq:ab_from_outcomes}
\end{equation}
The corresponding joint projector is
\begin{equation}
    \Pi_{r_1,r_2}
    =\frac{1}{4}(I+r_1C_1)(I+r_2C_2).
    \label{eq:joint_projector}
\end{equation}
Applied to $|++\rangle$, each nonzero branch is the desired state up to a Pauli frame,
\begin{equation}
    \frac{\Pi_{r_1,r_2}|++\rangle}
    {\|\Pi_{r_1,r_2}|++\rangle\|}
    =e^{i\phi_{ab}}Z_A^a Z_B^b|CS\rangle,
    \label{eq:branch_state}
\end{equation}
where $\phi_{ab}$ is an irrelevant global phase. Therefore no ideal logical branch must be discarded: the pair $(a,b)$ is retained as classical side information and absorbed into the Pauli frame of the delivered resource.

The four ideal branches are not equiprobable for the $|++\rangle$ input. Their probabilities are
\begin{equation}
    P_{00}=\frac{5}{8},\qquad
    P_{01}=P_{10}=P_{11}=\frac{1}{8}.
    \label{eq:branch_probabilities}
\end{equation}
This asymmetry changes only how often the branches occur; all four remain usable. Each joint eigenspace of $C_1$ and $C_2$ is one-dimensional. Once the two logical measurements are made, the state in each branch is therefore fixed to the Pauli-frame image in Eq.~\eqref{eq:branch_state}, rather than remaining an arbitrary logical state. We use this rank-one projection property when combining preparation, verification, and error detection.

\subsubsection{Logical Record Encoding}
\label{subsec:record_code}

Repeated non-Pauli checks are useful only if faults in the measurement record cannot easily transform one valid branch label into another. We therefore treat the logical outcomes themselves as a classical codeword. Define
\begin{equation}
    C_{12}\equiv C_1C_2.
    \label{eq:C12}
\end{equation}
Because $C_1$ and $C_2$ commute, an ideal branch labelled by $(a,b)$ has binary outcomes $a$, $b$, and $a\oplus b$ for $C_1$, $C_2$, and $C_{12}$, respectively.

A minimal five-check construction is
\begin{equation}
    C_1,\;C_2
    \;\big|\;D_{\mathrm{mid}}\;\big|\;
    C_{12},\;C_1,\;C_2,
    \label{eq:five_check_sequence}
\end{equation}
where $D_{\mathrm{mid}}$ denotes the midpoint physical error-detection stage described in the following section. Its ideal logical record is
\begin{equation}
    g_5(a,b)=(a,b,a\oplus b,a,b),
    \label{eq:g5}
\end{equation}
with codebook
\begin{equation}
    \mathcal C_5=
    \{00000,\,01101,\,10110,\,11011\}.
    \label{eq:C5_codebook}
\end{equation}
The minimum Hamming distance is $d_{\rm rec}=3$. Length five is the shortest binary code with four codewords and minimum distance three: the radius-one Hamming bound excludes length four, since $4(1+4)>2^4$, while Eq.~\eqref{eq:C5_codebook} gives an explicit length-five construction. We therefore use the five-check scheme as the minimal baseline.

For the performance protocol we instead use the symmetric six-check sequence
\begin{equation}
    C_1,\;C_2,\;C_{12}
    \;\big|\;D_{\mathrm{mid}}\;\big|\;
    C_1,\;C_2,\;C_{12},
    \label{eq:six_check_sequence}
\end{equation}
which produces
\begin{equation}
    g_6(a,b)=(a,b,a\oplus b,a,b,a\oplus b).
    \label{eq:g6}
\end{equation}
The four valid records are
\begin{equation}
    \mathcal C_6=
    \{000000,\,011011,\,101101,\,110110\}.
    \label{eq:C6_codebook}
\end{equation}
Every nonzero pairwise difference has Hamming weight four, giving
\begin{equation}
    d_{\rm rec}=4.
    \label{eq:drec4}
\end{equation}
The six-bit length is also minimal for this distance. Across four binary codewords there are six unordered pairs. A single binary coordinate can separate at most four of those pairs, while pairwise distance at least four requires a total pair-distance sum of at least $6\times4=24$. Hence a record of length $n$ obeys $4n\ge24$, or $n\ge6$. The sequence in Eq.~\eqref{eq:six_check_sequence} saturates this bound. The sixth check is therefore not an arbitrary additional repetition; it is the shortest binary encoding of the four Pauli-frame branches with distance four.

\begin{figure}[t]
    \centering
    \includegraphics[width=\textwidth]{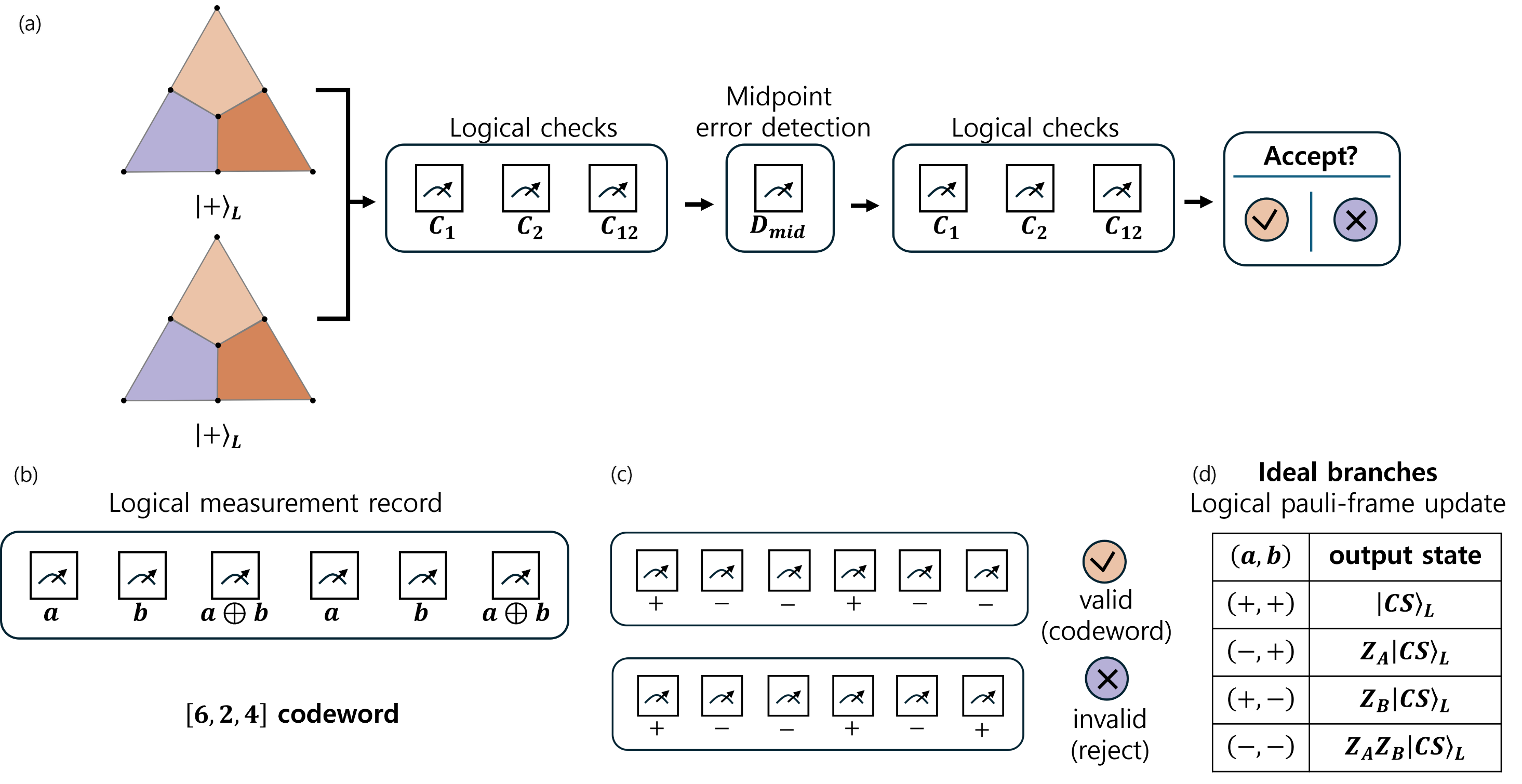}
    \caption{\textbf{Six-check measurement sequence and coded record.}
    (a) The two Steane blocks undergo $C_1$, $C_2$, and $C_{12}$ measurements, midpoint error detection $D_{\mathrm{mid}}$, and a second measurement triple.
    (b) The ideal history $g_6(a,b)=(a,b,a\oplus b,a,b,a\oplus b)$ forms a $[6,2,4]$ record code.
    (c) Records outside the four valid codewords are rejected.
    (d) All four ideal branches are usable and differ only by the Pauli-frame update $Z_A^aZ_B^b$.}
    \label{fig:six_check_record}
\end{figure}

The improvement is particularly relevant because a logical record bit is obtained from a seven-qubit cat-state readout. If each physical readout flips independently with probability $p_m$, the induced parity-flip probability of one logical record bit is
\begin{equation}
    q=\frac{1-(1-2p_m)^7}{2}.
    \label{eq:q_cat_parity}
\end{equation}
For the five-check code, a corrupted record is accepted as a different valid codeword with probability
\begin{align}
    P_{\mathrm{wrong},5}
    &=2q^3(1-q)^2+q^4(1-q) \\
    &=686p_m^3+O(p_m^4),
    \label{eq:pwrong5}
\end{align}
whereas the six-check code gives
\begin{align}
    P_{\mathrm{wrong},6}
    &=3q^4(1-q)^2 \\
    &=7203p_m^4+O(p_m^5).
    \label{eq:pwrong6}
\end{align}
At $p_m=10^{-3}$ these expressions give approximately $6.67\times10^{-7}$ and $6.93\times10^{-9}$, a factor of about 96.1 reduction in valid-to-valid record errors. This does not make the complete factory fourth order. It removes a large cubic record-aliasing contribution so that physical data faults and decoder errors set the leading accepted logical-error behavior.
The complete six-check measurement and record logic is summarized in Fig.~\ref{fig:six_check_record}.

\subsubsection{Steane-Code Implementation}
\label{subsec:steane_checks}

The Steane $[[7,1,3]]$ code is a canonical CSS/stabilizer code and provides the small-code host for our cultivation stage~\cite{Steane1996,CalderbankEtAl1997,Gottesman1997}. We cultivate the two logical qubits in a pair of Steane blocks. We use the logical convention
\begin{equation}
    X_L=X^{\otimes7},\qquad Z_L=Z^{\otimes7}.
    \label{eq:steane_logicals}
\end{equation}
For the standard codeword phase convention,
\begin{equation}
    S^{\otimes7}=S_L^\dagger,
    \qquad
    (S^\dagger)^{\otimes7}=S_L.
    \label{eq:steane_S_convention}
\end{equation}
Together with transversal $CZ$ between the two Steane blocks, this fixes the physical orientation of the coordinate checks. On coordinate $j\in\{1,\ldots,7\}$ we use
\begin{align}
    c_1^{(j)}
    &=X_{A_j}S_{B_j}^\dagger CZ_{A_jB_j},
    \label{eq:physical_coordinate_checks}\\
    c_2^{(j)}
    &=S_{A_j}^\dagger X_{B_j}CZ_{A_jB_j}.
    \label{eq:c2_coord}
\end{align}
Their transversal products realize the logical operators in Eqs.~\eqref{eq:C1}--\eqref{eq:C2}. The product check simplifies locally to
\begin{equation}
    c_{12}^{(j)}=c_1^{(j)}c_2^{(j)}
    =H_-^{(A_j)}H_-^{(B_j)},
    \qquad
    H_-\equiv\frac{X-Y}{\sqrt2}.
    \label{eq:c12_physical}
\end{equation}
The distinction between the logical $S$ orientation in Eqs.~\eqref{eq:C1}--\eqref{eq:C2} and the physical $S^\dagger$ factors in Eq.~\eqref{eq:physical_coordinate_checks} is essential; it follows directly from Eq.~\eqref{eq:steane_S_convention}.

\subsubsection{Controlled-Check Compilation}
\label{subsec:czz_compilation}

Multiqubit entangling gates have been demonstrated in superconducting circuits and Rydberg-atom arrays~\cite{Fedorov2012,Levine2019}. We do not use device-specific error models here. Instead, we reduce each coordinate check to a controlled application of $H\otimes H$. Define
\begin{equation}
    K=HS,
    \label{eq:K_def}
\end{equation}
where matrix products act from right to left. One exact choice of Clifford conjugations is
\begin{align}
    Q_1 &=(I\otimes K)\,CX_{A\rightarrow B}\,(K\otimes I),
    \label{eq:Qk}\\
    Q_2 &=(K\otimes I)\,CX_{B\rightarrow A}\,(I\otimes K),
    \label{eq:Q2}\\
    Q_{12}&=K\otimes K,
    \label{eq:Q12}
\end{align}
for which
\begin{equation}
    c_k=Q_k(H\otimes H)Q_k^\dagger,
    \qquad k\in\{1,2,12\}.
    \label{eq:QHHQ}
\end{equation}
The entangling $CX$ contained in $Q_1$ and $Q_2$ is necessary because $c_1$ and $c_2$ are themselves entangling two-qubit operators, whereas $c_{12}$ factorizes and requires no entangling basis change.

\begin{figure}[t]
    \centering
    \includegraphics[width=\textwidth]{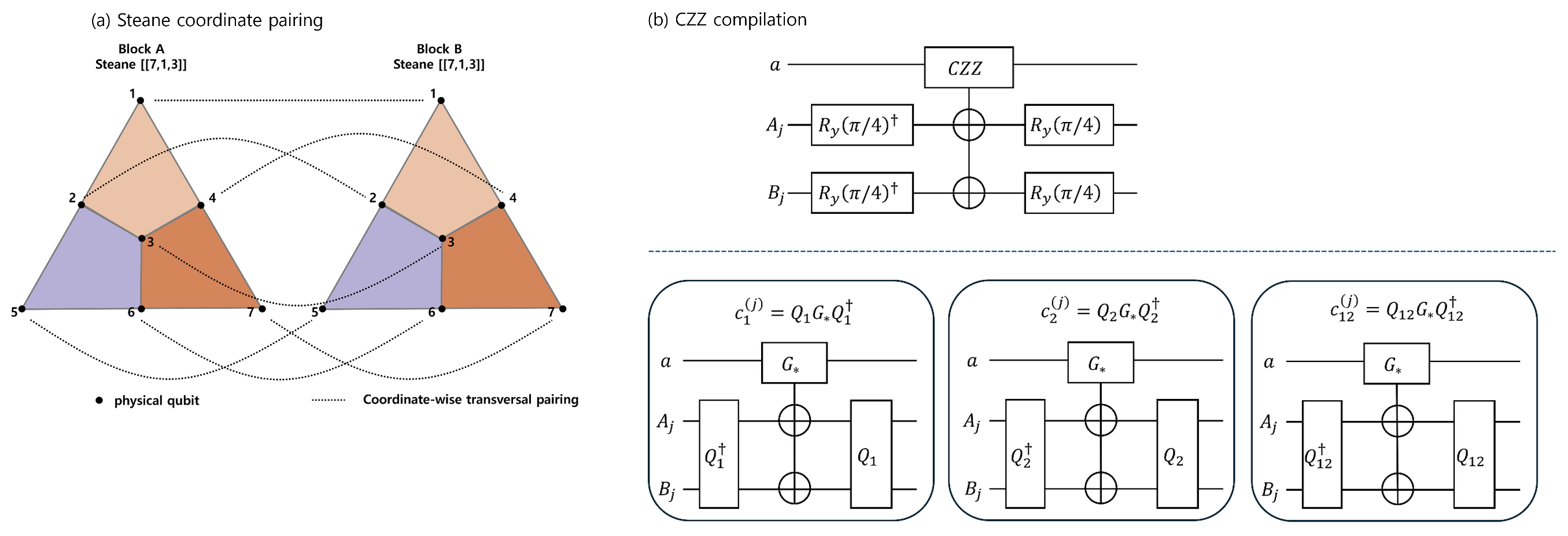}
    \caption{\textbf{Steane-coordinate checks and CZZ-based compilation.}
    (a) Matching coordinates of Steane blocks $A$ and $B$ realize the logical checks $C_1$, $C_2$, and $C_{12}$ through transversal products of the local operators.
    (b) A cat ancilla controls $G_*=C(H\otimes H)$, synthesized from one CZZ interaction and local $R_y(\pi/4)$ rotations. Clifford wrappers $Q_1$, $Q_2$, and $Q_{12}$ give the three coordinate checks. The six-check core contains $42$ CZZ, $56$ CX, and $224$ one-qubit locations.}
    \label{fig:czz_compilation}
\end{figure}

Let $a$ denote one cat ancilla coupled to a data-coordinate pair. The required controlled target primitive is
\begin{equation}
    G_*
    =|0\rangle\!\langle0|_a\otimes I_{AB}
    +|1\rangle\!\langle1|_a\otimes H_AH_B.
    \label{eq:Gstar}
\end{equation}
This primitive follows exactly from a controlled-$ZZ$ interaction. Writing
\begin{equation}
    \mathrm{CZZ}
    =|0\rangle\!\langle0|_a\otimes I_{AB}
    +|1\rangle\!\langle1|_a\otimes Z_AZ_B
    \label{eq:CZZ_def}
\end{equation}
and $R=R_y(\pi/4)$, for which $RZR^\dagger=H$, gives
\begin{equation}
    G_*=(I_a\otimes R_A R_B)\,
    \mathrm{CZZ}\,
    (I_a\otimes R_A^\dagger R_B^\dagger).
    \label{eq:CZZ_to_CHH}
\end{equation}
Combining Eqs.~\eqref{eq:QHHQ} and \eqref{eq:CZZ_to_CHH} gives a CZZ-based controlled implementation of all three coordinate checks. This is the physical primitive used in the cultivation circuit; we do not assume an equivalence to an iToffoli gate.
The coordinate pairing and controlled-check compilation are shown in Fig.~\ref{fig:czz_compilation}.

For the six-check sequence, the controlled-check core contains
\begin{equation}
    N_{\mathrm{check}}=42\ \mathrm{CZZ}+56\ CX+224\ (1Q)=322.
    \label{eq:check_core_inventory}
\end{equation}
This count includes only the check interactions and their basis-change wrappers. Cat-state preparation and readout, Steane preparation and error detection, and surface-code expansion are counted separately. Adjacent one-qubit factors compiled into the same physical pulse block are treated as one \emph{composite 1Q fault location}, with one stochastic fault opportunity. This convention gives the 224 one-qubit locations in Eq.~\eqref{eq:check_core_inventory}. The next section adds the remaining modules needed for the complete factory.

\subsection{Fault-Tolerant Modules and Code Expansion}
\label{sec:modules_expansion}

This section adds the fault-tolerant modules around the six-check core and the direct surface-code expansion that produces the final output patches. For each module, we state the low-order property needed by the end-to-end certification.

\subsubsection{Verified CAT$_7$ Readout}
\label{subsec:cat7}

Each coordinatewise realization of the logical checks described above is read out with a verified seven-qubit cat state~\cite{Shor1996,DiVincenzoShor1996}. One CAT$_7$ gadget spans all seven Steane coordinates of one logical-check measurement, so the six-check protocol uses exactly six verified CAT$_7$ gadgets in total, not one cat per coordinate. We use a balanced-tree CAT$_7$ preparation network together with a four-qubit verifier and partial-transversal verification pattern. The data cat is prepared through the tree
\begin{equation}
    0\!\rightarrow\!4,\qquad
    0\!\rightarrow\!2,\;4\!\rightarrow\!6,\qquad
    0\!\rightarrow\!1,\;2\!\rightarrow\!3,\;4\!\rightarrow\!5,
\end{equation}
while the verifier CAT$_4$ is prepared by
\begin{equation}
    7\!\rightarrow\!9,\qquad
    7\!\rightarrow\!8,\;9\!\rightarrow\!10.
\end{equation}
Verification is then performed by the partial-transversal pattern
\begin{equation}
    0\!\rightarrow\!7,\qquad
    2\!\rightarrow\!9,\qquad
    4\!\rightarrow\!8,\qquad
    5\!\rightarrow\!10,
\end{equation}
and the verifier qubits are measured in the $Z$ basis. We accept if and only if the four verifier outcomes are all equal, i.e., $0000$ or $1111$.

Explicit fault enumeration gives a simple support bound for accepted CAT$_7$ faults. Among 232 single-fault Pauli cases, 96 are accepted and none produces $X$ support greater than one on the data. Among 25,398 two-location Pauli cases, 5,524 are accepted and none produces $X$ support greater than two. The accepted gadget therefore satisfies
\begin{align}
    1\text{ fault} &\Longrightarrow \text{accepted }X\text{ support}\le 1, \\
    2\text{ faults} &\Longrightarrow \text{accepted }X\text{ support}\le 2.
\end{align}
These support bounds are the CAT$_7$ properties used in the later end-to-end analysis.

\subsubsection{State Preparation and Midpoint Error Detection}
\label{subsec:prep_mid_ed}

Each logical input block is prepared as a Steane-encoded $|+\rangle_L$ state and then checked immediately by full Steane error detection. The preparation starts from product states, applies an eight-CNOT encoder, and postselects on the error-detection result. The encoder uses the CNOT order
\begin{equation}
    0\!\rightarrow\!1,\;
    5\!\rightarrow\!3,\;
    6\!\rightarrow\!2,\;
    4\!\rightarrow\!1,\;
    0\!\rightarrow\!2,\;
    6\!\rightarrow\!3,\;
    5\!\rightarrow\!1,\;
    4\!\rightarrow\!6,
\end{equation}
with initial basis states
\begin{equation}
    q_0=|+\rangle,
    \qquad q_1,q_2,q_3=|0\rangle,
    \qquad q_4,q_5,q_6=|+\rangle.
\end{equation}
A separate logical-$X$ verification stage is not required in this architecture: detectable preparation faults are filtered by the immediate Steane error-detection step, while logical-normalizer preparation branches are subsequently projected by the rank-one logical checks described above.

For each block, the resulting preparation inventory is
\begin{equation}
    48\,CX + 14\,(1Q) + 21\,\text{prep} + 14\,M,
\end{equation}
for a total of 97 active locations per block and 194 active locations for the two-block input pair.

Midpoint Steane error detection is inserted between the two triples of logical checks. The routine measures the $Z$-type Steane syndromes with flagged ancillas~\cite{ChaoReichardt2018}, applies a transversal Hadamard, repeats the syndrome extraction for the complementary error class, and then applies the inverse transversal Hadamard. For one block this uses 40 CNOTs, 14 Hadamards, 14 ancilla preparations, and 14 measurements, for a total of 82 active locations.

Midpoint error detection keeps accepted residual errors at low weight between the two halves of the protocol. All 21 incoming weight-one Pauli errors are detected when the error-detection circuit is fault free. Among 698 single-fault cases, no accepted output has residual code weight above one; among 238,969 two-fault cases, none has residual code weight above two. The same weight-two bound holds for one incoming data Pauli combined with one error-detection fault. These are the residual-weight bounds used in the later low-order certificate.

\subsubsection{The Cultivation Core Cannot Be the Certified Output}
\label{subsec:why_escape}

A central architectural point is that the six-check Steane-level cultivation core is not, by itself, the certified factory output. Even though the coded measurement record removes the dominant readout-aliasing channel and the surrounding CAT and error-detection modules satisfy the required low-order fault bounds, late data faults after the final logical-check interactions can still evade the measurement record.

A representative example is the late two-fault pattern $Z_1Z_5$ on a Steane block. It has the same Steane syndrome as the single-qubit error $Z_3$. Minimum-weight correction therefore applies $Z_3$ and leaves $Z_1Z_3Z_5$, a zero-syndrome logical-$Z$ representative. The standalone Steane-level core can thus fail after two late faults. This failure is not specific to one decoder; it shows that the Steane-level state cannot serve as the certified factory output.

For this reason, the factory studied below includes the immediate transfer from the Steane core to larger surface-code patches. The certified circuit therefore consists of preparation, verified logical-check measurement, midpoint error detection, and direct code expansion.

\subsubsection{Direct $d=3\rightarrow 13$ Expansion}
\label{subsec:direct_expansion}

After the six-check protocol accepts, each Steane block is moved immediately into a distance-3 surface-code patch by lattice surgery~\cite{Horsman2012,Litinski2019Game} and then enlarged directly to distance 13. The Steane-to-$d=3$ escape uses 316 active locations per patch. The faultable $d=3\rightarrow13$ expansion round uses another 1,120, for a total of 1,436 active locations per patch. No additional distance-3 syndrome rounds are inserted between the escape and the expansion. The exact low-order audits below show that such rounds are not needed for the certificate. This downstream stage produces the surface-code state used as the operational output of the factory.

Table~\ref{tab:main_expansion_schedule} summarizes this schedule. The direct growth portion contains one faultable $d=3\rightarrow13$ expansion round. An ideal continuation is used only for later validation; it is not counted as an operational stage and never supplies information to the decoder.

\begin{table}[t]
\caption{Active-location inventory of the optimized downstream schedule, per output patch.}
\label{tab:main_expansion_schedule}
\begin{ruledtabular}
\begin{tabular}{l r}
Stage & Active locations \\
\hline
Steane-to-$d=3$ escape & 316 \\
Faultable $d=3\rightarrow13$ expansion & 1,120 \\
\quad Gate inventory: CX 624; M 84; MX 84 & \\
\quad Gate inventory: R 169; RX 159 & \\
\hline
Total per patch & 1,436 \\
\end{tabular}
\end{ruledtabular}
\end{table}

The complete factory contains $890+2\times1{,}436=3{,}762$ possible active locations. The earlier Direct schedule with a distance-5 escape, two distance-5 rounds, and staged $d=5\rightarrow7\rightarrow13$ growth contained 7,960, and an intermediate version with a single $d=5\rightarrow13$ round contained 5,154. The distance-3 interface is therefore 52.7\% and 27.0\% smaller, respectively. This schedule count is different from the accepted-output cost used later: rejected attempts can stop early, and several attempts may be needed before one output is accepted.

Supplementary Fig.~S2 (Supplementary Note~9) separates the 1,436-location downstream block into its escape and expansion parts and lists the gate inventory of the expansion round. The first-order mechanism counts in that figure count concrete Pauli mechanisms, not probability weights.

The expansion also defines the detector model used by the practical decoder. The following subsection, ``Decoder and Fault-Order Certification,'' describes this interface. The expansion is therefore part of the certified factory, not just a passive embedding of an already verified state.

\subsection{Decoder and Fault-Order Certification}
\label{sec:decoder_certification}

The direct expansion produces a correlated detector model. To avoid using information that would not be available when the state is delivered, we separate decoder inputs from later validation data. This section defines that separation, describes the practical decoder, and states the precise fault-order-three claim.

\subsubsection{Detector Information}
\label{subsec:truth_firewall}

For each output patch, the downstream schedule produces 304 detector bits. We partition these as
\begin{equation}
    304 = 136\;\text{operational} + 168\;\text{future-validation}.
    \label{eq:detector_partition}
\end{equation}
Only the first 136 bits are available to the acceptance and decoding rules. The remaining 168 future-validation bits are withheld until after the decoder has made its decision. They are used only to check whether the future boundary closes and whether a logical error remains after the decoded Pauli frame is applied.

The future validation continuation can reveal information that is unavailable when the state is delivered, so these 168 bits must not enter the decoder. In all simulations and audits we therefore enforce
\begin{equation}
    \mathcal D:\{0,1\}^{136}\longrightarrow
    \{I,X,Z,Y,\text{abstain}\},
    \label{eq:decoder_map}
\end{equation}
with the 168 future-validation bits excluded from the decoder input by construction.

The first-order effect catalogue provides an algebraic check of this information separation. For the decoder matrix, first-order fault mechanisms with the same detector pattern and logical effect are grouped into a common effect class. After restricting the correction model to the closed boundary, the decoder catalogue contains 1,068 such classes, including the identity. Removing the identity leaves 1,067 nontrivial columns. Let
\begin{equation}
    H\in\mathbb F_2^{136\times1067}
    \label{eq:H_matrix}
\end{equation}
be the operational detector-effect matrix and let
\begin{equation}
    H_{\rm aug}\in\mathbb F_2^{138\times1067}
    \label{eq:Haug_matrix}
\end{equation}
be the corresponding matrix augmented by the two logical-parity rows. The final matrix audit gives
\begin{equation}
    \operatorname{rank}H=136,
    \qquad
    \operatorname{rank}H_{\rm aug}=138,
    \label{eq:matrix_ranks}
\end{equation}
so the two logical rows add independent information beyond the operational detector span.

Before this restriction, the physical first-order model contains 1,071 effect classes. Three nonidentity classes have zero operational syndrome but a nonzero residual effect. They are excluded from the decoder matrix because they do not define valid correction columns at the operational boundary. They remain in the physical sampling and in the final validation, so they can still produce a scored failure when they occur.

\subsubsection{Low-Order Decoding}
\label{subsec:low_order_firewall}

The first decoder stage resolves all records covered by exact enumeration through total fault order two, including the allowed incoming boundary errors and downstream physical faults. Decisions made in this exact low-order stage are never replaced by a later approximate guess.

The order-two audit finds no closed-boundary logical failure in any required case. No single downstream effect with zero operational record causes a logical failure, and no pair of effects at distinct physical locations produces a failing zero-record event. Incoming boundary errors of weight one or two are also safe by themselves, and an incoming weight-one error combined with one downstream fault remains safe. A weight-two boundary error followed by one additional physical fault is already an order-three event and is therefore outside this low-order test.

If a record is resolved exactly at low order, the decoder keeps that certified Pauli frame. A record that cannot be resolved by the low-order model is next checked against a precomputed higher-order catalogue. Only records that remain unresolved are sent to the four-coset BP+OSD stage. This order preserves the exact low-order decisions while reserving approximate decoding for the harder records.

Finite-$p$ regression tests provide an independent implementation check. Across the ten-point sweep, no sampled closed-boundary logical failure of total fault order two or less is observed, and no decoding failure occurs among records sent to the higher-order BP+OSD stage. Across the sweep, 3,188 records reach the BP+OSD stage; every one has representatives in all four logical Pauli cosets, and none fails to decode.

\subsubsection{Four-Coset BP+OSD}
\label{subsec:bposd_decoder}

Records not resolved by the exact low-order stage are first checked against the precomputed higher-order catalogue. Catalogue matches require no further decoding. The remaining records are decoded with belief propagation plus ordered-statistics decoding (BP+OSD), evaluated separately in all four logical Pauli cosets. BP and OSD are established tools for sparse quantum-code decoding~\cite{PoulinChung2008,Roffe2020}. We use the augmented matrix $H_{\rm aug}$ so that detector consistency and logical-frame alternatives are treated in the same effect model. For each unresolved syndrome, the decoder requires representatives of all four logical Pauli cosets
\begin{equation}
    \{I,X,Z,Y\},
    \label{eq:four_cosets}
\end{equation}
so that all logical-frame alternatives are considered for each syndrome. For each unresolved record, the decoder searches the eight most plausible latent Steane sectors, ranked by detector mismatch and prior probability. Within each logical Pauli coset, candidates are combined into a log-sum-exp score. We define the decoder gap as the difference between the best and second-best coset scores; a larger gap indicates greater confidence in the selected logical frame.

The decoder architecture and all hyperparameters are fixed before the final performance evaluation. Table~\ref{tab:frozen_decoder_parameters_main} lists these settings.

\begin{table}[t]
\caption{Practical-decoder settings fixed for the final performance evaluation. Only the prior probabilities assigned to physical fault mechanisms are recalibrated when the operating error rate changes.}
\label{tab:frozen_decoder_parameters_main}
\begin{ruledtabular}
\begin{tabular}{l l}
Setting & Fixed value \\
\hline
Minimum decoder gap & 10 \\
Latent sectors searched & 8 \\
Maximum BP iterations & 30 \\
BP update rule & minimum-sum \\
Minimum-sum scaling & 0.625 \\
OSD method & OSD$_{\rm cs}$ \\
OSD order & 2 \\
Maximum effects per correction & 8 \\
\end{tabular}
\end{ruledtabular}
\end{table}
As $p$ changes, only the fault-mechanism prior is recalibrated; the decoder structure and hyperparameters remain fixed. The full $p$ sweep therefore uses one decoder design rather than a separately tuned decoder at each point.

The minimum decoder gap is the one setting that trades acceptance against logical quality, so we report two operating points. The default, $g=10$, is inherited from the earlier schedule and fixed before any evaluation. The optimized point is chosen on calibration data with seeds independent of all reported results: 10,000 finite-$p$ attempts and 101,864 order-three samples. The selection rule was fixed before the calibration outcomes were seen. It chooses the threshold in $\{4,\dots,10,12\}$ that minimizes the mean of $C_{\rm Direct}/C_{\rm 3T}$ at $p=5\times10^{-4}$ and $10^{-3}$, where the comparator is matched to the Direct $A_3^{\rm bin}(g)\,p^3$ target. This selects $g^\ast=4$. Changing the threshold cannot affect the low-order certificate: every record with a realization of total order at most two is resolved, or rejected as ambiguous, by the exact firewall before the BP+OSD stage is reached.

In summary, the decoder has three stages: exact low-order resolution, a precomputed higher-order catalogue, and four-coset BP+OSD. The fixed decoder-gap threshold then gives the final accept-or-abstain decision. Exact information is used whenever it is available, while BP+OSD is reserved for unresolved correlated records.

\subsubsection{Fault-Order-Three Certificate}
\label{subsec:fault_order_three}

We now define the logical-error event used in the rest of the paper. The decoder first processes only the 136 operational bits. If it accepts, the decoded Pauli frame is applied. The 168 future-validation bits are then used only for validation. An accepted sample with nonzero future syndrome is called a \emph{detectable residual}. If the future boundary closes but the final logical frame is wrong, we call the sample a \emph{closed-boundary logical failure}.

The exact structural audit and decoder regression find no accepted closed-boundary logical-failure mechanism through total fault order two. The order-three analysis, however, contains explicit accepted failure witnesses. Therefore the accepted closed-boundary logical-error rate has the expansion
\begin{equation}
    \mathrm{LER}(p)=A_3p^3+O(p^4)
    \label{eq:ler_fault_order_three}
\end{equation}
under the fixed practical decoder and the stated active-location stochastic-Pauli model.

We call this result \emph{fault order three}. It applies specifically to the accepted closed-boundary channel defined above, with the stated decoder, postselection rules, boundary condition, and noise model. It is not a general claim that the delivered state has conventional distance $d_f\ge3$. In particular, some operationally accepted outputs still contain residual errors that would be detected by the future validation continuation; we report those events separately.

The information flow is simple: only the 136 operational bits enter the decoder. They pass through exact low-order resolution, the higher-order catalogue, and, when needed, four-coset BP+OSD. The 168 future-validation bits are withheld until after the accept-or-abstain decision and are used only to score boundary closure and logical error. Exact enumeration excludes accepted logical-failure mechanisms through order two, while the order-three analysis supplies explicit failure witnesses. Together these results establish the leading $p^3$ behavior in Eq.~\eqref{eq:ler_fault_order_three} without using future-boundary information for decoding.

As a decoder diagnostic, we also perform a paired graphlike-PyMatching comparison on a dominant group of downstream order-three mechanisms. Minimum-weight perfect matching remains a central surface-code decoding paradigm~\cite{Fowler2015MWPM,HiggottGidney2025}, including recent extensions to decoding across transversal Clifford gates~\cite{SerraPeralta2026}. Because our plain graphlike construction drops irreducible detector hyperedges, we report the detailed comparison only in Supplementary Note~7 and do not use it to make a general claim about matching-based decoders.

\subsection{Finite-$p$ Factory Performance}
\label{sec:finite_p_performance}

\begin{figure}[t]
    \centering
    \includegraphics[width=0.92\textwidth]{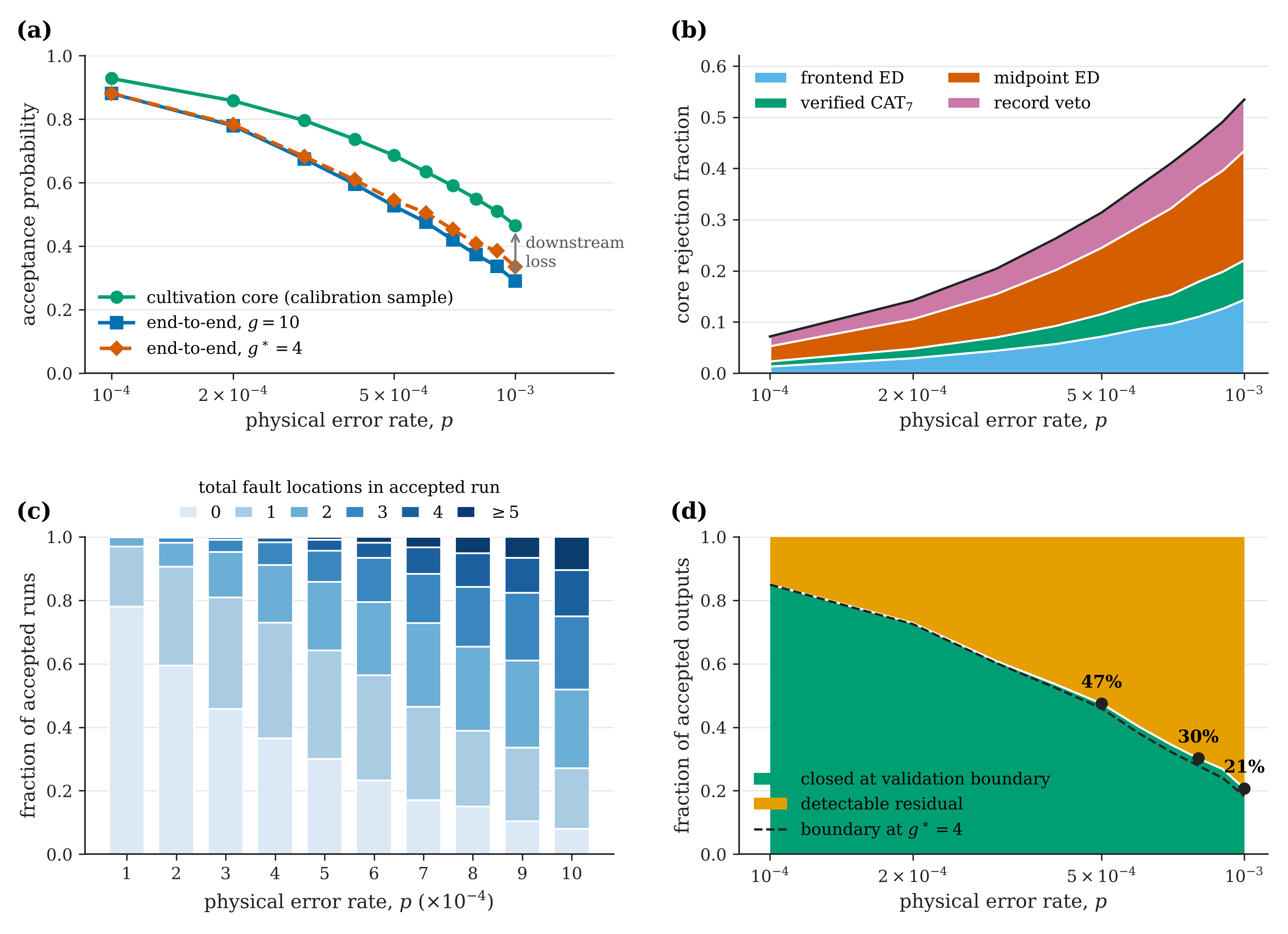}
    \caption{
Factory yield and accepted-output classification.
(a) Acceptance of the six-check core (calibration sample, 30,000 attempts per point) and of the complete factory (evaluation sample, 5,000 shots per point) at the default threshold $g=10$ and the calibrated threshold $g^\ast=4$.
(b) Sources of rejection before the downstream stage.
(c) Number of sampled fault locations in operationally accepted runs.
(c), (d) use the default threshold. (d) Accepted outputs are classified as either closed at the validation
boundary or containing a detectable residual syndrome; the dashed line marks the boundary at $g^\ast=4$.
Detectable residuals are not retrospectively rejected.
}
    \label{fig:factory_yield_anatomy}
\end{figure}

We evaluate the optimized Direct-$|CS\rangle$ factory at ten physical error rates from $10^{-4}$ to $10^{-3}$, using 5,000 independent shots at each point. The decoder structure and hyperparameters remain fixed; only the fault-mechanism prior is recalibrated for each $p$. Idle-memory faults are not included here; Sec.~\ref{sec:resource_comparison} treats them. Figure~\ref{fig:factory_yield_anatomy} shows the resulting acceptance and output classification.

\subsubsection{Acceptance and Postselection}

With the default threshold, the end-to-end operational acceptance falls from 0.881 at $p=10^{-4}$ to 0.527 at $5\times10^{-4}$, 0.373 at $8\times10^{-4}$, 0.336 at $9\times10^{-4}$, and 0.290 at $10^{-3}$. At $g^\ast=4$ the corresponding values are 0.882, 0.545, 0.408, 0.385, and 0.335. Figure~\ref{fig:factory_yield_anatomy}(b) shows where rejection occurs before the downstream stage. Midpoint Steane error detection accounts for about 40--41\% of these rejections across the sweep; frontend error detection, CAT verification, and the record veto make up the remainder. This breakdown explains the early-abort savings used in the resource analysis.

Figure~\ref{fig:factory_yield_anatomy}(c) provides a complementary view of postselection. The zero-fault fraction among operationally accepted samples falls rapidly with increasing $p$, while accepted runs containing several sampled fault locations become common. Operational acceptance therefore does not imply a fault-free state; it means only that the observed operational record satisfies the factory and decoder acceptance rules.

\subsubsection{Accepted-Output Validation}
\label{subsec:boundary_contract}

Operational acceptance does not guarantee that the future boundary will close. Among accepted outputs, the detectable-residual fraction rises from 15.0\% at $p=10^{-4}$ to 52.5\% at $5\times10^{-4}$, 69.8\% at $8\times10^{-4}$, 73.2\% at $9\times10^{-4}$, and 79.3\% at $10^{-3}$ [Fig.~\ref{fig:factory_yield_anatomy}(d)]. These outputs would produce a nontrivial syndrome in the future validation continuation, so we keep them separate from silent closed-boundary logical failures.

The validation step does not change the earlier accept-or-reject decision. Resource cost is therefore reported per operationally accepted output, and the 168 future-validation bits are used only to classify those outputs afterward. Dividing the cost by the fraction that closes at the future boundary would describe a different protocol that discards every detectable residual. We also do not charge the physical cost of an additional close-out QEC stage, either for Direct CS or for the three-$T$ comparator. Thus the two routes share the same no-extra-close-out-cost convention, although their underlying detector constructions are not identical. The resulting active-location metric is therefore narrower than a full spacetime-delivery cost.

\subsection{Leading-Order Logical Errors}
\label{sec:leading_order_decomposition}

The low-order certificate established above shows that closed-boundary logical failures begin at total fault order three. We therefore reconstruct two related leading coefficients,
\begin{align}
    P_{\rm fail}^{\rm bin}(p) &= A_3^{\rm bin}p^3+O(p^4),\\
    \mathcal I_{\rm closed}(p) &= A_3^{\rm infid}p^3+O(p^4),
\end{align}
Here $P_{\rm fail}^{\rm bin}$ counts whether a closed-boundary logical failure occurs, while $\mathcal I_{\rm closed}$ weights the same type of accepted output by state infidelity. These are different observables. Because the public three-$T$ Sinter data record binary logical errors, we use $A_3^{\rm bin}$ for the matched-LER comparison below and keep $A_3^{\rm infid}$ as a state-quality diagnostic.

\begin{figure}[t]
    \centering
    \includegraphics[width=0.92\textwidth]{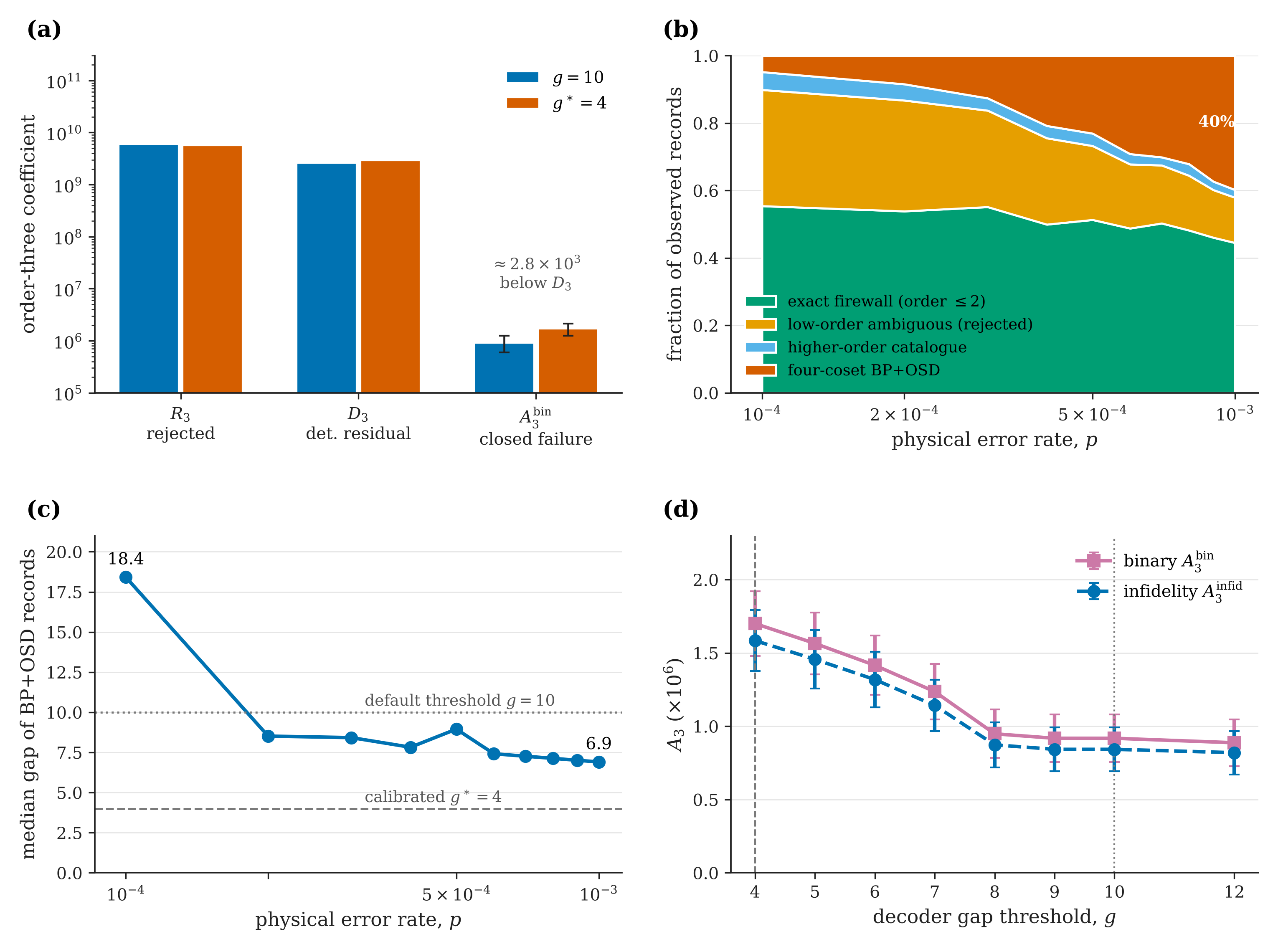}
    \caption{
Order-three error channels and decoder behavior.
(a) Leading coefficients at $p_{\rm dec}=5\times10^{-4}$ for
rejection $R_3$, detectable residuals $D_3$, and binary
closed-boundary logical failures $A_3^{\rm bin}$ (error bars $\pm2$ standard errors), at both thresholds.
Most order-three weight is rejected or remains detectable,
while silent logical failures are about $3\times10^3$ times rarer than detectable residuals.
(b) Observed nontrivial records resolved exactly by the order-two firewall, rejected as low-order ambiguous, resolved by the higher-order catalogue, or passed to four-coset BP+OSD.
(c) Median decoder gap of the records decoded by BP+OSD; the dotted and dashed lines mark the thresholds $g=10$ and $g^\ast=4$.
(d) Binary and infidelity-weighted $A_3$ as a function of the decoder-gap
threshold, from the same 300,001 order-three samples; points are therefore
correlated.
}
    \label{fig:order3_decoder_anatomy}
\end{figure}

\subsubsection{Order-Three Error Channels}

At $p_{\rm dec}=5\times10^{-4}$ and the default threshold, the reconstructed order-three coefficients are $R_3=6.06\times10^{9}$ for rejection, $D_3=2.60\times10^9$ for detectable residuals, and $A_3^{\rm bin}=9.2\times10^5$ for binary closed-boundary logical failures. Thus most third-order faults are either rejected or remain detectable at the future boundary. Only a small fraction become silent logical failures.

Figure~\ref{fig:order3_decoder_anatomy}(b,c) shows how decoder workload changes with $p$. The fraction of nontrivial records that reach the four-coset BP+OSD stage grows from 4.8\% at $10^{-4}$ to 39.8\% at $10^{-3}$, while the median gap falls from 18.4 to 6.9. Every audited record has full four-coset coverage and no BP+OSD decoding failure is observed. The trend therefore reflects increasing ambiguity at higher $p$.

\subsubsection{Order-Three Reconstruction}

We estimate the leading coefficients by stratified sampling conditioned on total fault order three. The 3,762 locations are partitioned into seven groups (front end, two CAT stages, two core stages, midpoint error detection, and downstream), which gives 84 strata. A 200-sample-per-stratum pilot fixes a Neyman allocation~\cite{Neyman1934}. A fresh sample of 300,001 order-three realizations is then drawn with independent seeds, so the estimator is unbiased. Every sample is scored at all candidate gap thresholds.

At the default threshold, $A_3^{\rm bin}=9.2(1.6)\times10^5$ and $A_3^{\rm infid}=8.4(1.5)\times10^5$, from 34 binary logical failures. At $g^\ast=4$, $A_3^{\rm bin}=1.70(0.22)\times10^6$ and $A_3^{\rm infid}=1.58(0.21)\times10^6$, from 63 failures. Of the failures in the default-threshold ensemble, 24 arise from three downstream faults, and the rest combine late core faults with downstream faults. Twelve of the thirteen recorded failure witnesses pass through the exact-firewall or structural-zero stages, which precede BP+OSD.

The same procedure applied to the distance-5 interface gives $A_3^{\rm bin}=1.87(0.36)\times10^6$, consistent with the earlier targeted estimate for that schedule. Like for like, the distance-3 interface therefore lowers $A_3^{\rm bin}$ by $51\pm13\%$, close to the 61\% reduction in the number of location triples. The estimates use the decoder prior at $5\times10^{-4}$. The 5,000-shot finite-$p$ sweep is used for acceptance and resource accounting, not for rare logical-error estimation.

\subsection{From Resource to Gate}
\label{sec:resource_to_gate}

The factory delivers the resource $Z_A^{a}Z_B^{b}|CS\rangle$ with a known branch label $r=(a,b)$. Here we show that standard gate teleportation~\cite{GottesmanChuang1999} consumes every branch with the usual Clifford feedback. Logical resource infidelity then maps directly to gate infidelity.

\paragraph{Lemma (known branch frames through ideal $CS$ injection).}
Let $D_A,D_B$ be data qubits, possibly entangled with a reference, and let $R_A,R_B$ hold the ideal resource $|\psi_r\rangle=Z_{R_A}^{a}Z_{R_B}^{b}|CS\rangle$ with $\mathrm{CS}=\operatorname{diag}(1,1,1,i)$. Apply $\mathrm{CNOT}_{D_A\to R_A}$ and $\mathrm{CNOT}_{D_B\to R_B}$, then measure $R_A,R_B$ in the $Z$ basis with outcome $m=(m_A,m_B)$. Define
\begin{equation}
U_m=\sum_{x,y=0}^{1} i^{(x\oplus m_A)(y\oplus m_B)}|xy\rangle\langle xy|,\qquad F_m=\mathrm{CS}\,U_m^\dagger .
\end{equation}
After the Clifford feedback $F_m$, the operation induced on the data is $\mathrm{CS}$ followed by the known Pauli frame $Z_{D_A}^{a}Z_{D_B}^{b}$. Every outcome occurs with probability $1/4$, independently of the data.

\paragraph{Proof.}
Write the data input as $\sum_{x,y}c_{xy}|xy\rangle$. The Kraus operator for outcome $m$ is
\begin{equation}
K_m^{r}=\frac12\sum_{x,y}(-1)^{a(x\oplus m_A)+b(y\oplus m_B)}\,i^{(x\oplus m_A)(y\oplus m_B)}|xy\rangle\langle xy|
=\frac{(-1)^{a m_A+b m_B}}{2}\,Z_{D_A}^{a}Z_{D_B}^{b}\,U_m .
\end{equation}
All factors are diagonal. Hence $F_mK_m^{r}=\tfrac12(-1)^{am_A+bm_B}Z_{D_A}^{a}Z_{D_B}^{b}\mathrm{CS}$ and $(K_m^{r})^\dagger K_m^{r}=I/4$. The identity extends to mixed inputs and arbitrary references by linearity. Up to global phase, the feedback is $F_{00}=I$, $F_{10}=CZ\,S_{D_B}^{\dagger}$, $F_{01}=CZ\,S_{D_A}^{\dagger}$, and $F_{11}=S_{D_A}S_{D_B}$; with the exact definition above, $F_{11}=-iS_{D_A}S_{D_B}$. $\square$

\paragraph{Gate fidelity.}
Let $\rho$ be any two-qubit resource state, and let $\mathcal E_r[\rho]$ be the injection channel above with all outcomes retained and the known branch $r$ compensated. Its entanglement fidelity with $\mathrm{CS}$ is
\begin{equation}
F_e\bigl(\mathrm{CS}^\dagger\circ\mathcal E_r[\rho]\bigr)=\langle\psi_r|\rho|\psi_r\rangle .
\end{equation}
For a pure resource with amplitudes $v_z$, the Kraus operator is $K_m=\sum_x v_{x\oplus m}|x\rangle\langle x|$. After feedback, $\operatorname{Tr}(\mathrm{CS}^\dagger Z^rF_mK_m)=(-1)^{r\cdot m}\sum_z(-1)^{r\cdot z}i^{-z_Az_B}v_z$. Then $F_e=\tfrac1{16}\sum_m|\operatorname{Tr}(\cdot)|^2$ gives the overlap. The mixed case follows by linearity.

\paragraph{Scope.}
The lemma keeps the factory's known $Z$ frame through ordinary Clifford feedback; it does not make that feedback Pauli. It does not extend to unknown errors: for example, $\mathrm{CS}\,X_{D_A}\mathrm{CS}^\dagger=X_{D_A}S_{D_B}CZ_{D_AD_B}$ is Clifford, not Pauli. It also does not include noisy feedback, recovery of detectable residuals, or buffering. Because the three-$T$ comparator below also produces a $|CS\rangle$ resource that is consumed in the same way, the consumption cost is common to both routes and is excluded from the comparison.

\subsection{Resource Comparison with Three-$T$ Synthesis}
\label{sec:resource_comparison}

\subsubsection{Accepted-Output Cost}

\begin{figure}[t]
    \centering
    \includegraphics[width=0.92\textwidth]{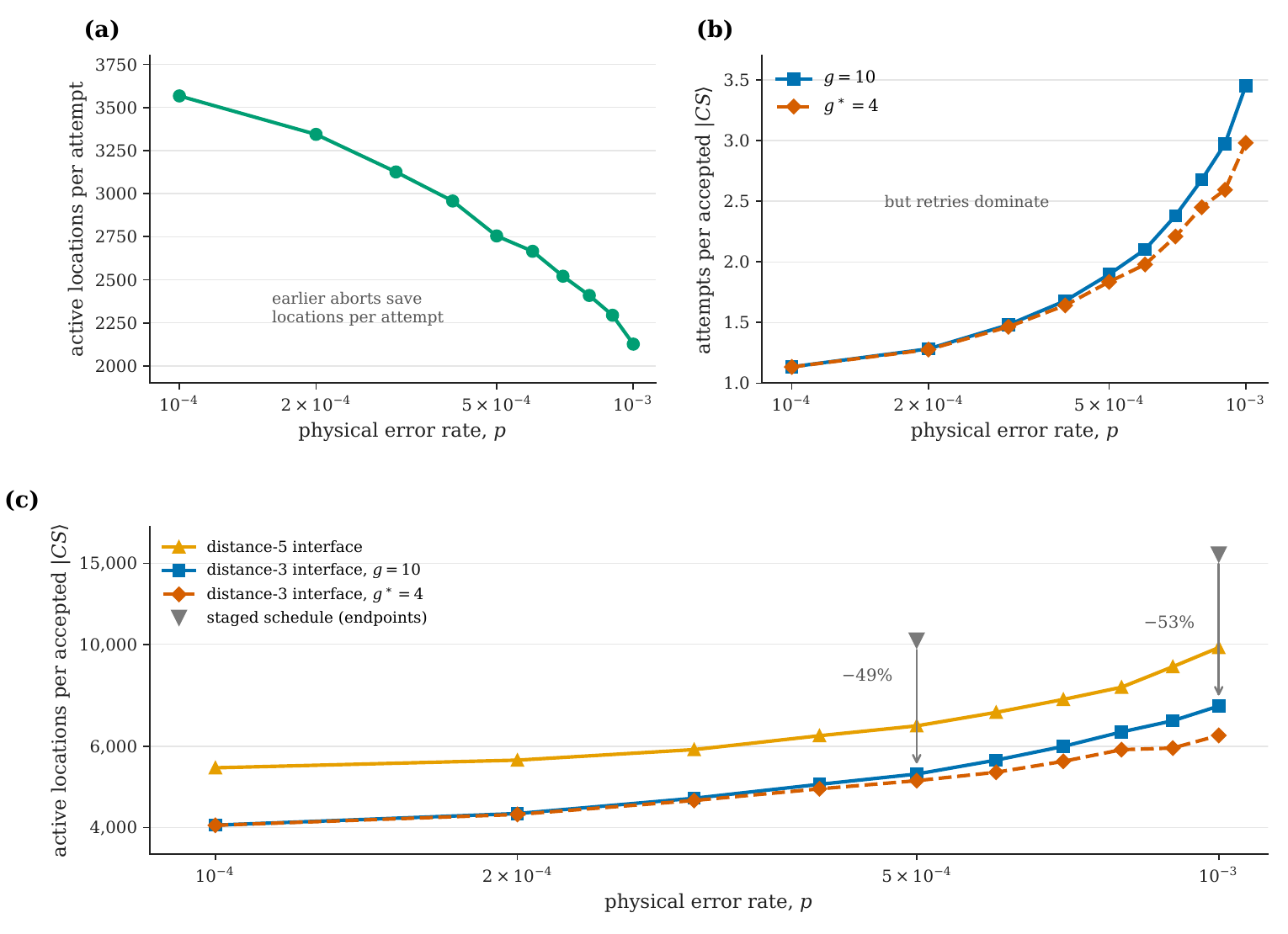}
    \caption{\textbf{Accepted-output resource cost.}
    (a) Mean active locations executed per attempt. Higher $p$ causes earlier aborts.
    (b) Expected attempts per accepted $|CS\rangle$ at both thresholds.
    (c) Active locations per operationally accepted $|CS\rangle$ for the distance-3 interface at both thresholds and for the distance-5 interface; for the earlier staged schedule only the two evaluated endpoints are shown, and arrows mark the reduction to the distance-3 interface at $g=10$.}
    \label{fig:resource_anatomy}
\end{figure}

For independent repeated attempts, the expected active-location cost per operationally accepted output is
\begin{equation}
    C_{\rm Direct}(p)=\frac{\overline N_{\rm act}^{\rm attempt}(p)}{P_{\rm acc}^{\rm op}(p)}.
\end{equation}
Figure~\ref{fig:resource_anatomy} shows the two competing effects entering this expression. As $p$ increases from $10^{-4}$ to $10^{-3}$, early rejection reduces the mean executed active locations per attempt from approximately 3,567 to 2,127, while the expected retry count rises from 1.13 to 3.45 at the default threshold (2.98 at $g^\ast$). The retry growth dominates, so accepted-output cost increases from approximately 4,048 to 7,344 active locations (6,341 at $g^\ast$).

At $p=5\times10^{-4}$, the distance-3 interface reduces the accepted-output cost from 10,203 for the earlier schedule, and from 6,651 for the single-round distance-5 interface, to 5,227. At $p=10^{-3}$ the corresponding values are 15,688, 9,836, and 7,344. These are within-architecture comparisons and do not depend on the three-$T$ baseline.

\subsubsection{Comparison with Three Cultivated $T$ States}

The indirect route prepares three cultivated $T$ states and converts them into $|CS\rangle$ with Eq.~\eqref{eq:cs_3t_identity_intro}. We compare the two routes on a common footing. Both use the same primitive convention: a reset followed by single-qubit gates is one preparation, and single-qubit gates followed by a measurement are one measurement. Both use the same noise model without idle-memory faults, the same early-abort accounting, and the same output boundary. The comparator is the public MSC circuit of Ref.~\cite{Sahay2026} at $d_{\rm final}=13$. One $T$ path costs 1,400 active locations: 236 in the $d=5$ cultivation stage, 816 in the $d=5\to7$ growth with three $d=7$ rounds, and 348 in the unitary $d=7\to13$ growth. The conversion to $|CS\rangle$ is charged as three transversal CX gates and one measurement at $d=13$, which is 676 locations.

We re-decoded the comparator rather than reuse the public Sinter rows, for two reasons. The public rows include idle-memory noise, which the Direct model omits. They also compute the acceptance gap with a final noiseless $d=13$ round, whereas Direct accepts using operational detectors only. In addition, rejected comparator attempts are charged only the locations executed before the abort, as for Direct, rather than the full path cost. We implemented a complementary-gap matching decoder~\cite{Bombin2024,GidneyShuttyJones2024,HiggottGidney2025} for the same circuit. It reproduces the public rows to within 0.5 percentage points in discard rate and 10--20\% in logical error rate, and it is slightly weaker than the reference decoder (Supplementary Note~8). Every three-$T$ value below uses this decoder. For each $p$, the comparator uses the cheapest gap setting whose logical error rate meets the Direct target $A_3^{\rm bin}p^3$. The comparator is therefore held to the quality of the Direct factory at each Direct threshold.

\begin{table}[t]
\caption{Active locations per accepted $|CS\rangle$ and spacetime ratio, Direct CS (distance-3 interface) against the re-decoded three-$T$ comparator matched to $A_3^{\rm bin}p^3$. Ratios below one favor Direct CS. ``Storage'' is synchronous assembly from three parallel factories. The spacetime ratio counts qubit-ticks from qubit lifetimes under one rule for both routes (Supplementary Note~8). Direct statistical errors are 1.3--2.2\%.}
\label{tab:resource_comparison}
\begin{ruledtabular}
\small
\begin{tabular}{c c r r r r r}
Threshold & $p$ & $C_{\rm Direct}$ & $C_{\rm 3T}$ & Ratio & Ratio, storage & Spacetime ratio \\
\hline
$g=10$ & $5\times10^{-4}$ & 5,227 & 5,167 & 1.01 & 0.49 & 0.63 \\
 & $8\times10^{-4}$ & 6,451 & 5,211 & 1.24 & 0.53 & 0.75 \\
 & $9\times10^{-4}$ & 6,824 & 5,153 & 1.32 & 0.59 & 0.80 \\
 & $1\times10^{-3}$ & 7,344 & 5,180 & 1.42 & 0.60 & 0.88 \\
$g^\ast=4$ & $5\times10^{-4}$ & 5,056 & 5,162 & 0.98 & 0.47 & 0.61 \\
 & $8\times10^{-4}$ & 5,904 & 5,075 & 1.16 & 0.57 & 0.70 \\
 & $9\times10^{-4}$ & 5,953 & 4,965 & 1.20 & 0.66 & 0.72 \\
 & $1\times10^{-3}$ & 6,341 & 4,969 & 1.28 & 0.67 & 0.79 \\
\end{tabular}
\end{ruledtabular}
\end{table}

Table~\ref{tab:resource_comparison} summarizes the comparison. Without any cost for holding $T$ states, Direct CS at $g^\ast$ uses 2\% fewer active locations than the comparator at $p=5\times10^{-4}$, a difference of about 1.6 Direct standard errors and thus consistent with parity, and 16--28\% more at $8\times10^{-4}$--$10^{-3}$. Without idle noise, the comparator meets its target with little gap postselection, so its cost is nearly independent of $p$. The Direct cost instead rises with $p$ because acceptance falls.

The cost decomposition at $p=0$ shows where the Direct overhead went (Supplementary Note~8). The Direct source core costs 890 locations against $3\times236=708$ for three $T$ cultivations. The escape costs $2\times316$, which the three-$T$ route does not need. Growth to $d=13$ costs $2\times1{,}120$ against $3\times1{,}164$, and Direct needs no conversion. The total is 3,762 against 4,876. The escape was 40\% of the zero-noise cost with a distance-5 interface and is 17\% with the distance-3 interface. The remaining Direct deficit at higher $p$ is therefore acceptance, not location count.

\begin{figure}[t]
    \centering
    \includegraphics[width=0.96\textwidth]{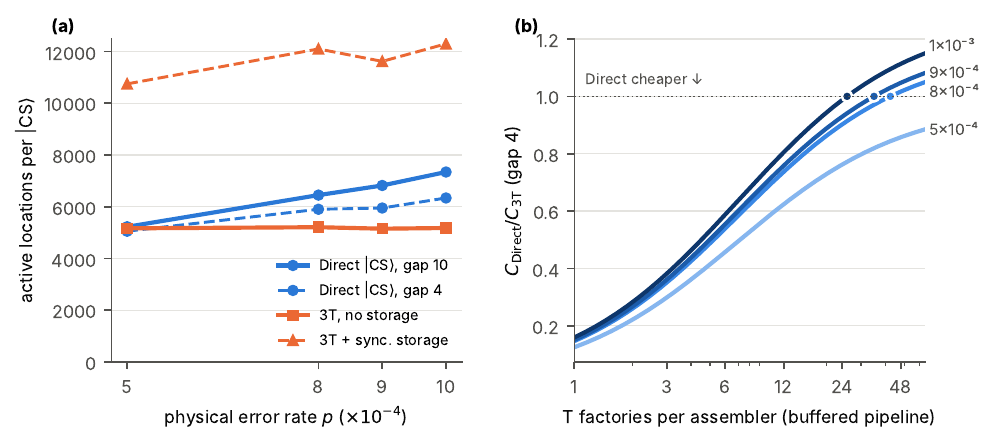}
    \caption{\textbf{Comparison with three cultivated $T$ states.}
    Same primitive convention, noise model without idle faults, early-abort accounting and output boundary for both routes; each three-$T$ point uses the cheapest gap setting meeting the Direct target $A_3^{\rm bin}p^3$.
    (a) Active locations per accepted $|CS\rangle$ for Direct CS at the default and calibrated decoder thresholds, and for the comparator matched to the default-threshold target. The comparator is shown without storage (solid) and with synchronous three-factory assembly (dashed).
    (b) Cost ratio at $g^\ast=4$ when $k$ factories feed one assembler through a buffer (mean holding time $3\bar t_T/k$ per $|CS\rangle$, $\bar t_T$ the time per accepted $T$). Dots mark the break-even factory count.}
    \label{fig:resource_comparison}
\end{figure}

The comparison above assigns no cost to holding $T$ states before conversion. Direct CS has no such cost, because both output patches are produced by a single accepted attempt. The three-$T$ route must hold earlier $T$ states as $d=13$ patches while it waits for the remaining ones, at 960 active locations per patch-round. At $g^\ast$, Direct CS is cheaper whenever this holding cost exceeds 0.9, 1.0, and 1.4 $d=13$ patch-rounds per $|CS\rangle$ at $p=8\times10^{-4}$, $9\times10^{-4}$, and $10^{-3}$. At $5\times10^{-4}$ it is cheaper with no holding cost at all. Synchronous assembly from three parallel factories exceeds these thresholds at every error rate, and Direct is then 33--53\% cheaper [Fig.~\ref{fig:resource_comparison}(a)]. With $k$ factories feeding a buffer, the holding cost falls as $1/k$. The comparator becomes cheaper beyond about 42, 35, and 25 factories per assembler at the three higher error rates [Fig.~\ref{fig:resource_comparison}(b)].

Active locations do not price how long qubits are held. We therefore also count qubit-ticks from the qubit lifetimes in the Stim~\cite{Gidney2021Stim} circuits of both routes, using one rule. The Direct core, which is not a Stim circuit, uses its native schedule; that rectangle bound overstates its lifetime count. Under this rule, Direct CS uses 21--39\% less spacetime than the comparator at $g^\ast$ (12--37\% at $g=10$), before any storage is charged. The Direct output patches are small for most of their lifetime, whereas the $T$ factory grows to distance 7 and holds three rounds there. The result depends on the counting convention. In the convention least favorable to Direct, the Direct downstream is counted as a rectangle of all touched qubits over all ticks, and the comparator uses its published successful-path volume with rejected attempts free. In that convention the spacetime ratio rises to 0.98--1.27 at $g^\ast$. We therefore regard spacetime parity or better as established at $p\lesssim5\times10^{-4}$, and the sign at $10^{-3}$ as convention-dependent.

\subsubsection{Resource Assumptions}

Four assumptions remain favorable to the three-$T$ route. First, the conversion is charged as three transversal CX gates and one measurement, with no syndrome rounds between them; lattice-surgery CX gates would cost of order $d$ rounds each~\cite{Horsman2012,Litinski2019Game}. Second, the conversion's own logical errors are neglected. Third, routing between the factories and the assembler is free. Fourth, rejected comparator attempts are charged up to the first postselected detector that fires, whereas a rejected Direct attempt is charged for the whole stage in which it is rejected (front end, core with midpoint error detection, or downstream); the Direct ledger is therefore the coarser one.

Three choices remain favorable to Direct CS. First, $A_3^{\rm bin}$ counts only closed-boundary logical failures, so residuals that the withheld validation round would detect are not failures. Applying the same definition to the three-$T$ comparator shows that it rewards withholding information. In the public noise model at $p=10^{-3}$, the comparator's closed logical error rate is $1.8\times10^{-3}$ without an extra noisy $d=13$ round before the withheld round and $3.6\times10^{-3}$ with it, because more failures become detectable residuals. Second, a CZZ gate is counted as one location. Third, the comparator is simulated with a separate noise location for each Hadamard, although its cost is counted in the merged convention.

\subsubsection{Idle-Memory Noise}

The comparison above omits idle-memory faults for both routes. We add them with one rule: in every layer, a qubit that holds a state but is not acted on receives single-qubit depolarizing noise of strength $p$. Idle counts depend on the schedule, and the Direct front end and core had no layer schedule. We therefore schedule both routes with the same procedure. Each qubit keeps its order of operations, and the layer of every operation is chosen to minimize the total number of idle locations. All constraints are differences of two layer indices, so the linear program has an integral optimum. Because the operations on each qubit are unchanged, the certificates above still apply. The schedule has 207 idle locations in the Direct front end and core, 233 per output patch, and 1,000 per $T$ path. The public comparator schedule has 1,951 per $T$ path under the same rule. The idle locations are simulated explicitly. They add no joint effect class to the decoder model, and no sampled closed-boundary failure of total fault order at most two occurs (Supplementary Note~10).

Idle noise lowers the Direct acceptance at $g^\ast$ from 0.545 to 0.441 at $p=5\times10^{-4}$ and from 0.335 to 0.229 at $10^{-3}$. Its cost per accepted output rises by 15\% and 24\%, to 5,796 and 7,874 active locations. The comparator cost rises by 4\% and 8\%, because it is dominated by location count rather than acceptance. Without storage, the ratio becomes 1.08 and 1.46, against 0.98 and 1.28 without idle noise. The order-three coefficient with idle noise was not re-estimated. The comparator target is scaled by $k=1.6$, the growth of the dominant location triples; for $k$ from 1 to 2.5 the ratio ranges over 1.04--1.09 and 1.36--1.54 (Supplementary Note~10). Direct CS remains cheaper whenever the comparator holds more than 0.4 and 2.6 distance-13 patch-rounds per $|CS\rangle$ at the two error rates, and under synchronous three-factory assembly the ratios are 0.37 and 0.42. Idle noise therefore narrows the regime without storage in which Direct CS is competitive, but leaves the storage argument intact. All operations, including measurement and reset, occupy one layer in this model. Slower measurement would raise the idle load of both routes, and that of Direct CS more, because its core contains more mid-circuit measurements.

Leakage, coherent faults, and correlated faults are omitted for both routes.

\section{Discussion}
\label{sec:discussion}

Cultivation of entangled resources has precedent in the logical $|CX\rangle$ protocol~\cite{Vaknin2026}. Concurrent work cultivates logical $|CCZ\rangle$ states directly on bivariate tricycle and bicycle codes, using their native cup-product $CCZ$ gates and decoder-based postselection~\cite{HirasakiLee2026}. Both works show that a multiqubit non-Clifford resource can be cultivated without first preparing $T$ states. They differ in the code family, the check mechanism, and the output interface: the present construction uses joint Clifford checks on small codes, a protected branch record, and a surface-code output with an exact low-order certificate. Our contribution is a direct construction for the specific $|CS\rangle$ resource. It combines rank-one joint logical projections, a protected $[6,2,4]$ branch record, and an explicitly audited code-growth boundary. All four ideal logical branches can be used with known Pauli-frame updates (Sec.~\ref{sec:resource_to_gate}). The low-order certificate concerns the accepted channel at the closed validation boundary under the stated noise model and fixed decoder; it does not certify a noisy logical consumer.

The logical distinction from $CX$ projection is useful independently of resource costs. For two logical qubits, $(I+CX)/2$ has rank three, whereas the four joint projectors of $C_1$ and $C_2$ each have rank one. Repeated projection onto a degenerate eigenspace does not by itself select a unique state. In the cited $CX$ construction, initialization selects the resource, and two copies implement a Toffoli with probability $4/9$~\cite{Dennis2001,Vaknin2026}. Our ideal $CS$ injection instead uses every measurement outcome after standard Clifford feedback. Because the target operations differ, this contrast is not a quantitative cost advantage over the $CX$ protocol.

Specialized multiqubit distillation is complementary to direct cultivation. Gong et al.\ construct compact $CS$-input protocols, including a four-to-one $CS$ protocol, and consider cultivated input resources in their benchmarks~\cite{Gong2026}. A symmetry-assisted search has also found distillation factories with entangled outputs that combine $T$, $CS$, and $CCZ$ states~\cite{JainWillsSingh2026}. They note that cultivating $CS$ states had not been studied directly, and the present construction supplies a concrete candidate for that input. Establishing the performance of a combined cultivation-and-distillation pipeline would additionally require its input-error correlations and resource accounting.

The coded record is a particularly portable design principle. Six bits are the minimum length for placing four binary messages at pairwise Hamming distance four, and the construction changes the record-only valid-to-valid error channel from cubic to quartic order in the physical CAT-readout flip rate. This does not make the complete factory fourth order. Instead, it removes a large classical-record aliasing channel and exposes the remaining physical and decoding mechanisms that determine the third-order accepted logical channel.

The order-three channel decomposition gives a simple message. Rejection and detectable residuals carry several orders of magnitude more third-order weight than silent closed-boundary logical failures. In addition, more records require four-coset BP+OSD as $p$ increases. These trends point to postselection, downstream syndrome structure, and correlated decoding as the main targets for further improvement.

The large detectable-residual fraction at the upper end of the simulated range makes the output boundary important. Acceptance is decided without the withheld validation bits, so a residual detected later is still counted as an operationally accepted output. Such an event is not a silent closed-boundary logical failure; it means that the withheld continuation would reveal a nontrivial syndrome. The future validation continuation is also not charged as an extra physical QEC stage. In Sec.~\ref{sec:resource_comparison} the three-$T$ comparator is evaluated with the same convention: its final noiseless round is withheld from acceptance and charged no cost. Applied to that comparator, the closed-boundary definition rewards withholding information, so we report the comparison under three output conventions (Supplementary Note~8). This intentionally operational metric describes delivery before an additional close-out QEC stage; a full hardware-level resource model should add that stage together with common timing assumptions to both routes.

The controlled checks are compiled through a CZZ-based realization of $C(H\otimes H)$. Algebraically, the CZZ of Eq.~\eqref{eq:CZZ_def} is exactly $CZ_{aA}CZ_{aB}$, so a native three-body gate is not required at the unitary level. The present resource and noise model nevertheless treats CZZ as one three-qubit fault location; replacing it by two physical CZ locations would change both the location count and the fault distribution and would require a separate performance audit. Multiqubit entangling operations have been demonstrated in superconducting and neutral-atom platforms~\cite{Fedorov2012,Levine2019}. Our parameter $p$ is a uniform active-location stochastic-Pauli scale and should not be identified directly with a measured hardware gate error.

The present model omits leakage, coherent and calibration errors, correlated CZZ faults across coordinates, routing, and operation durations. The three-$T$ comparator is simulated with the same omissions. Idle-memory faults are treated in Sec.~\ref{sec:resource_comparison} under one rule and one optimized schedule for both routes. They affect Direct CS more because its cost is set by acceptance. The decisive open item is therefore a timing model in which measurement and reset are slower than gates. The scheduling program extends to it directly by weighting each constraint with the duration of the earlier operation.

Finally, the leading logical coefficients come from targeted sampling of exactly three faults. The 5,000-shot finite-$p$ sweep was designed for acceptance and resource accounting, not for rare logical-error scaling. A future higher-statistics finite-$p$ study could independently cross-check the reconstructed $p^3$ behavior without changing the fixed architecture.

Taken together, these results establish a composed direct-cultivation route for the entangled non-Clifford resource $|CS\rangle=CS|++\rangle$: a protected branch record, verified small-code modules, immediate code growth, an operational/validation detector contract, and a practical decoder are treated as one composed factory object. Within the stated active-location stochastic-Pauli model and fixed decoder, accepted closed-boundary logical failures are excluded through total fault order two and explicit order-three mechanisms remain. The escape-and-growth interface is the decisive cost term. A distance-3 escape followed by one $d=3\rightarrow13$ growth round keeps the certificate, halves the leading logical-error coefficient relative to a distance-5 interface, and lowers the accepted-output cost by about half relative to the earlier staged schedule. Against three cultivated $T$ states matched to the same binary-LER target and charged nothing for $T$-state storage, Direct CS uses 2\% fewer active locations at $p=5\times10^{-4}$ and 16--28\% more at $8\times10^{-4}$--$10^{-3}$. It needs no $T$-state storage or assembly, and under a single qubit-tick rule for both routes it uses less spacetime at every error rate studied, although the sign at $10^{-3}$ depends on the counting convention. Idle-memory noise widens the active-location gap without storage to 8\% and 46\%, while Direct CS remains cheaper once $T$ states must be held for more than 0.4 and 2.6 patch-rounds. The central result is architectural: an entangled non-Clifford state can itself be cultivated with a protected branch history and an explicitly certified output interface.

\section{Methods}
\label{sec:methods}

\subsection{Simulation and noise model}
The numerical analysis uses the optimized Direct-$|CS\rangle$ architecture, the fixed practical decoder, and the structural audits described in the Results. We use an active-location stochastic-Pauli model with a common physical error scale $p$ applied to the component fault rules. Fault order counts distinct active physical locations. Idle-memory errors are not included in the main results; Supplementary Note~10 adds them to both routes under a common rule and an idle-minimizing schedule that keeps every qubit's order of operations. Adjacent one-qubit factors compiled into the same physical pulse block are treated as one composite 1Q fault location and receive one stochastic fault opportunity. The parameter $p$ is therefore a simulation scale and is not identified with a hardware-specific randomized-benchmarking error rate. Full module inventories and low-order enumeration details are given in Supplementary Notes~1--4.

\subsection{Operational boundary and finite-$p$ sampling}
The output interface contains 136 operational detector bits per patch and 168 future-validation bits. Only the operational bits are available to acceptance and decoding. The future-validation bits are withheld until after a decoding decision and are then used to distinguish detectable residuals from closed-boundary logical failures. At each of ten physical-error points from $10^{-4}$ to $10^{-3}$, 5,000 independent evaluation shots are generated after fresh prior calibration. Rejected attempts can stop early, and detectable residuals are not retrospectively rejected. The ideal validation continuation is not charged as an additional physical close-out QEC stage.

\subsection{Practical decoder and low-order certification}
The practical decoder first resolves records covered by exact enumeration through total fault order two, then queries a precomputed higher-order catalogue, and finally applies four-coset BP+OSD to unresolved records. Exact low-order decisions are never overwritten by a later approximate stage. The fixed decoder settings are a minimum gap of 10 (default) or 4 (selected on independent calibration data), eight latent sectors, at most 30 minimum-sum BP iterations with scaling 0.625, OSD$_{\rm cs}$ of order 2, and at most eight effects per correction. Only the physical-mechanism priors are recalibrated as $p$ changes. The low-order audit checks downstream single faults and distinct-location pairs together with the allowed incoming boundary-error cases and finds no accepted closed-boundary logical failure through total fault order two. Detailed matrix and decoder-validation audits are given in Supplementary Notes~4 and~5.

\subsection{Order-three reconstruction}
The leading closed-boundary error coefficient is estimated by stratified sampling conditioned on total fault order three, over 84 strata of seven location groups. A pilot fixes a Neyman allocation; a fresh sample of 300,001 realizations with independent seeds is then scored at every candidate gap threshold. Standard errors are estimator standard errors. Supplementary Note~6 gives the details.

\subsection{Resource accounting and three-$T$ comparator}
The Direct resource metric is the expected number of active physical locations executed per operationally accepted $|CS\rangle$, including early-abort savings and retry overhead. The indirect comparator is the public $d_{\rm final}=13$ MSC circuit of Ref.~\cite{Sahay2026}. It is recounted in the Direct primitive convention, at 1,400 active locations per $T$ path, and followed by a 676-location conversion. We re-decoded it in Stim~\cite{Gidney2021Stim} with a complementary-gap PyMatching decoder~\cite{HiggottGidney2025}, with idle-memory faults removed and the final noiseless round withheld from acceptance. Rejected attempts are charged up to the first firing postselected detector. For each $p$, we select the cheapest gap setting whose three-$T$ logical error rate meets $A_3^{\rm bin}p^3$. $T$-state storage is charged at 960 active locations per distance-13 patch-round. Spacetime is counted in qubit-ticks from qubit lifetimes in the Stim circuits of both routes, with the Direct core from its native schedule; Supplementary Note~8 gives the decoder validation, the three output conventions, the storage and spacetime models, and the zero-noise cost decomposition.

\section*{Acknowledgments}
This work was supported by the IITP (Institute of Information \& Communications Technology Planning \& Evaluation)-ITRC (Information Technology Research Center) grant funded by the Korea government (Ministry of Science and ICT) (IITP-2026-RS-2021-II211810). This work was supported by the National Research Foundation of Korea (NRF) grant funded by the Korea government (MSIT) (No.~RS-2023-00242396). This research was supported by the Quantum Science and Technology Flagship Project (Quantum Computing) (No.~RS-2025-25463854) through the National Research Foundation of Korea (NRF) funded by the Korean government (Ministry of Science and ICT (MSIT)).

\section*{Data availability}
The numerical data, structural certificates, analysis outputs, and reproducibility materials supporting this study are publicly available in the Direct-CS repository~\cite{MinDirectCS2026}.

\section*{Code availability}
The custom simulation, decoding, fault-enumeration, and resource-analysis code supporting this study is publicly available in the Direct-CS repository~\cite{MinDirectCS2026}.

\section*{Author contributions}
G.M. conceived and conducted the study, performed all theoretical and numerical analyses, developed the software, analyzed the results, prepared the figures, and wrote the manuscript. J.H. advised the research.

\section*{Competing interests}
The authors declare no competing interests.

\bibliographystyle{unsrtnat}
\bibliography{direct_cs_references}

\end{document}


\begin{titlepage}
\centering
\vspace*{0.22\textheight}
{\LARGE\bfseries Supplementary Information\par}
\vspace{1.2em}
{\Large Direct Cultivation of Entangled $|CS\rangle$ Magic States\par}
\vspace{1.2em}
{\large Gunsik Min and Jun Heo\par}
\vspace{0.6em}
School of Electrical Engineering, Korea University, Seoul 02841, Republic of Korea
\vfill
\end{titlepage}
\section{Supplementary Note 1: Simulation and Resource Conventions}
\label{supp:scope}

The numerical results use the optimized Direct-$|CS\rangle$ architecture, the fixed practical decoder, and the structural and leading-order audits described below.

The stochastic model is an active-location Pauli noise model with a common physical error scale $p$ applied to the component fault rules. Fault order counts distinct active physical locations, and idle-memory errors are not included for either the Direct factory or the three-$T$ comparator except in Supplementary Note~10. Adjacent one-qubit factors compiled into one physical pulse block are one composite 1Q fault location and receive one stochastic fault opportunity. The manuscript does not identify $p$ with a hardware-specific randomized-benchmarking error rate.

The resource metric is the expected number of active physical locations actually executed per \emph{operationally accepted} $|CS\rangle$. Rejected attempts can stop early. Acceptance is decided without the 168 future-validation bits; those bits are used only afterward to identify detectable residuals and closed-boundary logical failures. Detectable residuals are not retrospectively rejected, and the ideal validation continuation is not charged as a physical close-out QEC stage. The three-$T$ comparator is evaluated with the same boundary convention (Supplementary Note~8). This metric is therefore narrower than a full delivered-state spacetime cost.

Table~\ref{tab:full_inventory} lists the full optimized potential-location budget. The six CAT$_7$ entries correspond to one verified seven-coordinate cat gadget for each of the six logical checks.

\begin{table}[H]
\caption{Full-potential active-location inventory of the Direct-$|CS\rangle$ factory with the distance-3 interface.}
\label{tab:full_inventory}
\begin{ruledtabular}
\begin{tabular}{l r}
Component & Active locations \\
\hline
Two Steane preparations + initial ED & 194 \\
Six-check controlled-check core & 322 \\
Midpoint Steane ED, two blocks & 164 \\
Six verified CAT$_7$ gadgets & 210 \\
\hline
Pre-downstream subtotal & 890 \\
Downstream patch A & 1,436 \\
Downstream patch B & 1,436 \\
\hline
Full potential total & 3,762 \\
\end{tabular}
\end{ruledtabular}
\end{table}

The earlier staged schedule contains 7,960 and the single-round distance-5 interface 5,154 full-potential active locations, so the distance-3 interface reduces the structural count by 52.7\% and 27.0\%, respectively. This structural count is distinct from accepted-output cost, which additionally reflects early abort and retries.

\section{Supplementary Note 2: Logical-Record Code}
\label{supp:record}

The two logical measurement outcomes are represented by bits
\begin{equation}
 a=(1-r_1)/2,\qquad b=(1-r_2)/2,
\end{equation}
where $r_1,r_2\in\{+1,-1\}$ are the eigenvalues of the commuting logical involutions $C_1$ and $C_2$. The main six-check record is
\begin{equation}
 g_6(a,b)=(a,b,a\oplus b,a,b,a\oplus b),
\end{equation}
while the minimal five-check baseline is
\begin{equation}
 g_5(a,b)=(a,b,a\oplus b,a,b).
\end{equation}
Table~\ref{tab:record_codewords} lists the four ideal records.

\begin{table}[H]
\caption{Ideal codewords of the five- and six-check logical records. The six-check record has pairwise Hamming distance four, while the five-check baseline has minimum distance three.}
\label{tab:record_codewords}
\begin{ruledtabular}
\begin{tabular}{c c c c}
$a$ & $b$ & $g_5(a,b)$ & $g_6(a,b)$ \\
\hline
0 & 0 & 00000 & 000000 \\
0 & 1 & 01101 & 011011 \\
1 & 0 & 10110 & 101101 \\
1 & 1 & 11011 & 110110 \\
\end{tabular}
\end{ruledtabular}
\end{table}

If a CAT$_7$ parity readout is formed from seven independent physical measurement bits with flip probability $p_m$, the parity-flip probability is
\begin{equation}
 q=\frac{1-(1-2p_m)^7}{2}.
\end{equation}
For the five-check record, the probability that measurement noise maps an ideal codeword into a \emph{different valid codeword} is
\begin{align}
 P_{\mathrm{wrong},5}
 &=2q^3(1-q)^2+q^4(1-q) \\
 &=686p_m^3+O(p_m^4).
\end{align}
For the six-check record,
\begin{align}
 P_{\mathrm{wrong},6}
 &=3q^4(1-q)^2 \\
 &=7203p_m^4+O(p_m^5).
\end{align}
At $p_m=10^{-3}$ these expressions give approximately $6.67\times10^{-7}$ and $6.93\times10^{-9}$, respectively, a record-only reduction by a factor of approximately 96.1. This comparison concerns the reliability of the classical logical record only and is not a full-factory resource or logical-error improvement factor.

A length-five binary representation of four messages cannot place all six message pairs at Hamming distance at least four. Each coordinate contributes to at most four of the six pairwise distances, so five coordinates contribute at most 20 in total, whereas six pairs each at distance at least four would require a total distance of at least 24. Therefore six bits are the minimum length for this four-word distance-four record.

\section{Supplementary Note 3: Module-Level Certificates}
\label{supp:module_certificates}

\subsection{Verified CAT$_7$ gadget}

The CAT$_7$ data ancilla is prepared with the balanced tree
\begin{equation}
0\!\to\!4,\quad
0\!\to\!2,\;4\!\to\!6,\quad
0\!\to\!1,\;2\!\to\!3,\;4\!\to\!5,
\end{equation}
and the CAT$_4$ verifier is prepared with
\begin{equation}
7\!\to\!9,\qquad 7\!\to\!8,\;9\!\to\!10.
\end{equation}
Partial-transversal verification uses
\begin{equation}
0\!\to\!7,\quad2\!\to\!9,\quad4\!\to\!8,\quad5\!\to\!10,
\end{equation}
followed by $Z$ measurements of the four verifier qubits. The gadget accepts only when the four outcomes are all equal. The explicit fault enumeration is summarized in Table~\ref{tab:cat_cert}.

\begin{table}[H]
\caption{Verified CAT$_7$ low-order support certificate.}
\label{tab:cat_cert}
\begin{ruledtabular}
\begin{tabular}{l r r r}
Fault set & Enumerated & Accepted & Accepted violations \\
\hline
Single physical Pauli fault & 232 & 96 & $X$ support $>1$: 0 \\
Two-location Pauli faults & 25,398 & 5,524 & $X$ support $>2$: 0 \\
\end{tabular}
\end{ruledtabular}
\end{table}

Thus one accepted fault produces $X$ support at most one and two accepted faults produce $X$ support at most two.

\subsection{Midpoint Steane error detection}

The midpoint Steane error-detection gadget uses the three $Z$-syndrome supports $0145$, $0235$, and $0246$. A transversal Hadamard is then applied, the same syndrome-extraction structure is repeated, and a transversal Hadamard returns the block to its original basis. Per block the gadget uses 40 CNOTs, 14 Hadamards, 14 ancilla preparations, and 14 measurements, for 82 active locations. Table~\ref{tab:ed_cert} summarizes the explicit low-order checks.

\begin{table}[H]
\caption{Midpoint Steane error-detection certificate.}
\label{tab:ed_cert}
\begin{ruledtabular}
\begin{tabular}{l r l}
Audit & Cases & Result \\
\hline
Incoming weight-one $X/Y/Z$ & 21 & all detected fault-free \\
Single ED Pauli fault & 698 & accepted residual code weight $>1$: 0 \\
Two ED faults & 238,969 & accepted residual code weight $>2$: 0 \\
Incoming single Pauli + one ED fault & -- & accepted residual code weight $>2$: 0 \\
\end{tabular}
\end{ruledtabular}
\end{table}

\subsection{Physical $|+\rangle_L$ preparation}

Each Steane block is prepared with an eight-CNOT encoder followed immediately by full Steane error detection. The encoder CNOT order is
\begin{equation}
0\!\to\!1,
5\!\to\!3,
6\!\to\!2,
4\!\to\!1,
0\!\to\!2,
6\!\to\!3,
5\!\to\!1,
4\!\to\!6,
\end{equation}
with $q_0,q_4,q_5,q_6$ initialized in $|+\rangle$ and $q_1,q_2,q_3$ initialized in $|0\rangle$. The total inventory is 97 active locations per block. The preparation audit contains 127 single-encoder-fault classes: 117 reject, and the ten accepted survivors are logical normalizers ($I$: 3, $X_L$: 7). Among 7,161 two-encoder-fault cases, 6,938 reject and the 223 accepted survivors are also normalizers ($I$: 111, $X_L$: 97, $Z_L$: 8, $Y_L$: 7). No detectable survivor remains in either audit.

For one encoder fault combined with one initial-error-detection fault, 88,646 combinations were enumerated and 86,550 rejected. The remaining two-fault cases can contain normalizer or detectable residual components; the two-fault budget is then exhausted, and the later fault-free midpoint error-detection stage rejects detectable residuals while logical normalizers are projected by the rank-one logical-check structure.

\section{Supplementary Note 4: Expansion and Detector-Model Audit}
\label{supp:firewall}

The downstream schedule contains 1,436 active locations per patch:
\begin{itemize}
\item 316 for the Steane-to-$d=3$ lattice-surgery escape;
\item 1,120 for one faultable direct $d=3\rightarrow13$ expansion round, made of 624 CX, 84 M, 84 MX, 169 R, and 159 RX locations.
\end{itemize}
No intermediate distance-3 syndrome round is inserted. The ideal follow-up used for validation adds no operational fault locations and supplies no decoder information.

The circuit is generated by the same Steane--surface lattice-surgery generator and expansion module used for the distance-5 interface. Only the escape distance changes, and the logical frame is re-derived exactly. Both the X- and Z-basis clean splices are deterministic; this is checked by Stim~\cite{Gidney2021Stim} detector-error-model construction and noiseless sampling. Table~\ref{tab:supp_expansion_schedule} lists the schedule.

\begin{table}[H]
\caption{Downstream schedule inventory, per patch.}
\label{tab:supp_expansion_schedule}
\begin{ruledtabular}
\begin{tabular}{l r}
Stage / gate type & Active locations \\
\hline
Steane-to-$d=3$ escape & 316 \\
Direct $d=3\rightarrow13$ faultable round & 1,120 \\
\quad CX & 624 \\
\quad M & 84 \\
\quad MX & 84 \\
\quad R & 169 \\
\quad RX & 159 \\
\hline
Total per patch & 1,436 \\
\end{tabular}
\end{ruledtabular}
\end{table}

\emph{Structural audit.} For each X- or Z-recovery probe, the audit represents 11,728 concrete physical Pauli mechanisms: 4,912 from the escape and 6,816 from the expansion round. After joint X/Z merging, the complete set contains 12,734 mechanisms, grouped into 3,178 effect classes. Using all detectors, no single physical effect and no pair at distinct physical locations produces an accepted logical failure with a zero record. Canonical incoming Steane sectors of weight one or two are safe alone, and a weight-one sector combined with one physical fault is also safe.

\emph{Operational audit.} The downstream detector record contains 304 bits: 136 operational bits and 168 future-validation bits used only for validation. The exact firewall over the 136 operational bits joins every truth-closed realization of total order at most two. These realizations comprise incoming canonical sectors of weight 0, 1, or 2, combined with zero, one, or two downstream faults at distinct locations, respecting the order budget. The firewall contains 418,224 records. Of these, 402,661 have a unique logical Pauli and 15,563 are ambiguous and therefore rejected. A zero operational record is compatible only with the identity through total order two.

The physical first-order catalogue contains 1,071 effect classes. Three nonidentity classes have zero operational syndrome and are not valid correction columns at the closed boundary. They are excluded from the decoder matrix but remain in physical sampling and final validation. The resulting decoder model contains 1,068 effect classes, or 1,067 nontrivial columns. The matrix audit gives
\begin{equation}
H\in\mathbb{F}_2^{136\times1067},\quad \operatorname{rank}(H)=136,\qquad
H_{\rm aug}\in\mathbb{F}_2^{138\times1067},\quad \operatorname{rank}(H_{\rm aug})=138.
\end{equation}

\begin{table}[H]
\caption{Operational-detector and matrix audit for the distance-3 interface. Values for the earlier distance-5 interface are given for reference.}
\label{tab:matrix_audit}
\begin{ruledtabular}
\begin{tabular}{l r r}
Quantity & $d=3$ interface & $d=5$ interface \\
\hline
Physical active locations per patch & 1,436 & 2,132 \\
Operational detector bits & 136 & 263 \\
Future-validation bits & 168 & 168 \\
Future-validation bits decoder-visible & No & No \\
Concrete mechanisms (joint X/Z) & 12,734 & 19,590 \\
Raw first-order effect classes & 1,071 & 3,343 \\
Nonidentity classes with zero operational syndrome & 3 & 3 \\
Effect columns & 1,067 & 3,339 \\
$\operatorname{rank}(H)$ / $\operatorname{rank}(H_{\rm aug})$ & 136 / 138 & 263 / 265 \\
Firewall records (ambiguous) & 418,224 (15,563) & 4,818,356 (14,197) \\
Zero-record low-order gate & PASS & PASS \\
\end{tabular}
\end{ruledtabular}
\end{table}

We also audited, with the same code, the distance-5 interface with one or no distance-5 wait rounds (2,004 and 1,876 locations per patch) and the distance-3 interface with one or two distance-3 wait rounds (1,476 and 1,516). All pass both gates. Wait rounds are therefore not required by the certificate. In the finite-$p$ evaluation they did not improve acceptance enough to pay for their cost.

\section{Supplementary Note 5: Decoder Validation}
\label{supp:decoder}

The practical decoder has three stages: exact low-order resolution, a precomputed higher-order catalogue, and four-coset BP+OSD for the remaining records. Table~\ref{tab:decoder_params} gives the settings. Only the minimum gap differs between the two reported operating points.

\begin{table}[H]
\caption{Practical-decoder parameters.}
\label{tab:decoder_params}
\begin{ruledtabular}
\begin{tabular}{l l}
Parameter & Value \\
\hline
Minimum decoder gap & 10 (default); 4 (calibrated, $g^\ast$) \\
Latent sectors searched & 8 \\
BP maximum iterations & 30 \\
BP rule & minimum-sum \\
Minimum-sum scaling & 0.625 \\
OSD method & OSD$_{\rm cs}$ \\
OSD order & 2 \\
Maximum effects per correction & 8 \\
\end{tabular}
\end{ruledtabular}
\end{table}

At each finite-$p$ operating point, 5,000 evaluation shots are generated after a fresh prior calibration with 30,000 shots. Table~\ref{tab:decoder_audit_10p} lists the observed operational records and the subset that reaches the four-coset BP+OSD stage. Every such record has four-coset coverage, and no decoding failure is observed. No sampled closed-boundary logical failure of total fault order two or less occurs at any point or threshold.

\begin{table}[H]
\caption{Practical-decoder audit across the ten finite-$p$ operating points. ``BP+OSD records'' remain unresolved after the exact low-order and catalogue stages.}
\label{tab:decoder_audit_10p}
\begin{ruledtabular}
\small
\begin{tabular}{c r r c c c c}
$p$ & Records & BP+OSD & Failures & Median gap & Acc.\ ($g=10$) & Acc.\ ($g^\ast$) \\
\hline
$1\times10^{-4}$ & 415 & 20 & 0 & 18.42 & 0.881 & 0.882 \\
$2\times10^{-4}$ & 581 & 49 & 0 & 8.52 & 0.779 & 0.783 \\
$3\times10^{-4}$ & 769 & 97 & 0 & 8.42 & 0.675 & 0.683 \\
$4\times10^{-4}$ & 957 & 199 & 0 & 7.82 & 0.596 & 0.609 \\
$5\times10^{-4}$ & 1,058 & 244 & 0 & 8.95 & 0.527 & 0.545 \\
$6\times10^{-4}$ & 1,273 & 371 & 0 & 7.42 & 0.476 & 0.505 \\
$7\times10^{-4}$ & 1,418 & 427 & 0 & 7.26 & 0.420 & 0.453 \\
$8\times10^{-4}$ & 1,540 & 495 & 0 & 7.13 & 0.373 & 0.408 \\
$9\times10^{-4}$ & 1,635 & 610 & 0 & 7.01 & 0.336 & 0.385 \\
$1\times10^{-3}$ & 1,699 & 676 & 0 & 6.91 & 0.290 & 0.335 \\
\end{tabular}
\end{ruledtabular}
\end{table}

\emph{Exact ILP decoding.} The operational matrix is small enough for exact minimum-weight decoding. We therefore augmented the BP+OSD candidate list in each coset with the exact integer-linear-programming optimum and two further solutions, using the same parity priors. On paired finite-$p$ samples, no decision changed. BP+OSD already reaches the relevant coset optima, and the remaining abstentions reflect genuine near-ties.

\section{Supplementary Note 6: Order-Three Reconstruction}
\label{supp:a3}

The fault-order statement applies to the accepted closed-boundary logical-error channel under the practical decoder and the stated noise and boundary model. No accepted logical-failure mechanism occurs through total fault order two, while explicit order-three failure witnesses are present.

We retain two third-order output-quality observables. The binary coefficient $A_3^{\rm bin}$ counts whether an accepted sample that closes at the future boundary has a logical failure; rejection and detectable residuals contribute zero. The infidelity-weighted coefficient $A_3^{\rm infid}$ weights the same class of closed samples by state infidelity. The three-$T$ comparison uses the binary coefficient as its quality target (Supplementary Note~8).

\emph{Estimator.} The 3,762 locations form seven groups: front end, CAT before the midpoint, core before the midpoint, midpoint error detection, CAT after the midpoint, core after the midpoint, and downstream. Together they define 84 order-three strata covering all $\binom{3762}{3}=8.87\times10^9$ location triples. A pilot with 200 samples per stratum fixes a Neyman allocation~\cite{Neyman1934} $n_s\propto T_s\sqrt{\hat q_s(1-\hat q_s)}$, with $T_s$ the triple mass and $\hat q_s$ a Jeffreys-smoothed pilot failure rate, and a floor of 200. A fresh sample of 300,001 realizations is then drawn with independent seeds; the pilot enters only through the allocation. Each realization is decoded once with the decoder prior at $5\times10^{-4}$ and scored at every candidate threshold.

\begin{table}[H]
\caption{Order-three coefficients versus the decoder-gap threshold, from 300,001 realizations (reporting ensemble). $R_3$ and $D_3$ are in units of $10^9$, $A_3$ in units of $10^6$; errors are estimator standard errors. The last column is the independent calibration ensemble (101,864 realizations) used only for threshold selection.}
\label{tab:a3_final}
\begin{ruledtabular}
\small
\begin{tabular}{c r c c c c c}
Gap & Failures & $A_3^{\rm bin}$ & $A_3^{\rm infid}$ & $R_3$ & $D_3$ & $A_3^{\rm bin}$ (calibration) \\
\hline
4 & 63 & $1.70\pm0.22$ & $1.58\pm0.21$ & $5.69$ & $2.94$ & $1.82\pm0.40$ \\
5 & 58 & $1.57\pm0.21$ & $1.46\pm0.20$ & $5.69$ & $2.94$ & $1.73\pm0.39$ \\
6 & 52 & $1.42\pm0.20$ & $1.32\pm0.19$ & $5.70$ & $2.93$ & $1.59\pm0.38$ \\
7 & 45 & $1.24\pm0.19$ & $1.14\pm0.18$ & $5.77$ & $2.87$ & $1.59\pm0.38$ \\
8 & 35 & $0.95\pm0.17$ & $0.87\pm0.15$ & $6.00$ & $2.66$ & $1.32\pm0.34$ \\
9 & 34 & $0.92\pm0.16$ & $0.84\pm0.15$ & $6.02$ & $2.63$ & $1.32\pm0.34$ \\
10 & 34 & $0.92\pm0.16$ & $0.84\pm0.15$ & $6.06$ & $2.60$ & $1.32\pm0.34$ \\
12 & 33 & $0.89\pm0.16$ & $0.82\pm0.15$ & $6.07$ & $2.59$ & $1.32\pm0.34$ \\
\end{tabular}
\end{ruledtabular}
\end{table}

At the default threshold, 24 of the 34 failures come from three downstream faults. Seven come from two late core faults with one downstream fault, and three from one late core fault with two downstream faults. Twelve of the thirteen recorded witnesses pass through the exact-firewall or structural-zero stages, which precede BP+OSD.

\emph{Reference value for the distance-5 interface.} The identical procedure, with the same seed and the same allocation rule, gives $A_3^{\rm bin}=1.87(0.36)\times10^6$ and $A_3^{\rm infid}=1.67(0.33)\times10^6$ from 29 failures. The earlier targeted estimate for that schedule was $2.37(0.50)\times10^6$, so the two are consistent. The distance-3 interface therefore lowers $A_3^{\rm bin}$ by $51\pm13\%$ at the default threshold.

\section{Supplementary Note 7: PyMatching Comparison}
\label{supp:mwpm}

This diagnostic was performed on the distance-5 interface and is retained as a statement about plain graphlike matching on these detector models. We evaluated a plain graphlike PyMatching decoder at $p_{\rm dec}=2\times10^{-4}$ on one dominant group of downstream order-three mechanisms, keeping the same 10,000 physical samples and BP+OSD accept-or-reject decisions and changing only the Pauli frame assigned to accepted high-order records. The BP+OSD frame gave $A_3^{\rm BP}=8.396(3.462)\times10^5$; the graphlike-MWPM reframe gave $A_3^{\rm MWPM}=3.410(0.231)\times10^7$, a ratio of about 40.6 with paired $z\simeq14.5$. The graphlike reduction does not represent the irreducible detector hyperedges and detector-free logical-only mechanisms of the audited models, so this result applies only to that approximation.

\section{Supplementary Note 8: Three-$T$ Comparison}
\label{supp:3t}

\subsection{Comparator and primitive convention}

The indirect comparator prepares three cultivated $T$ states and converts them into a $|CS\rangle$ resource using
\begin{equation}
CS=(T_A\otimes T_B)\,\mathrm{CNOT}_{A\rightarrow B}
(I\otimes T_B^\dagger)\,\mathrm{CNOT}_{A\rightarrow B}.
\end{equation}
Each $T$ state comes from the public fold-transversal MSC circuit of Ref.~\cite{Sahay2026} with final distance 13. The conversion is charged as three transversal CX gates and one measurement at distance 13, $4\times169=676$ active locations. No syndrome rounds are inserted between the conversion gates, and conversion errors are neglected. Both choices favor the comparator.

The comparator circuit counts each Hadamard as a separate location. The Direct factory instead uses the native convention of Supplementary Note~1: a reset followed by single-qubit gates is one preparation (RX), single-qubit gates followed by a measurement are one measurement (MX), and any other run of adjacent single-qubit gates is one composite location. Table~\ref{tab:3t_convention} recounts the public circuit in both conventions. The explicit count reproduces the earlier figure of 1,662 locations per path exactly. The Direct downstream circuit contains 241 RX, 252 MX, and no Hadamards, so both conventions give it the same count.

\begin{table}[H]
\caption{Active locations of one three-$T$ preparation path in the two primitive conventions.}
\label{tab:3t_convention}
\begin{ruledtabular}
\begin{tabular}{l r r}
Stage & Explicit & Native (Direct convention) \\
\hline
$d=5$ cultivation & 283 & 236 \\
$d=5\rightarrow7$ growth with three $d=7$ rounds & 971 & 816 \\
Unitary $d=7\rightarrow13$ growth & 408 & 348 \\
\hline
Total per $T$ path & 1,662 & 1,400 \\
\end{tabular}
\end{ruledtabular}
\end{table}

\subsection{Re-decoding the comparator}

The public Sinter rows differ from the Direct evaluation in three ways:
\begin{enumerate}
\item They include depolarizing idle-memory noise, which the Direct model omits. The noisy part of the circuit has 1,912 idle-noise locations.
\item The acceptance gap uses a final noiseless distance-13 round. Direct acceptance uses operational detectors only.
\item Each rejected attempt is charged the full path cost, whereas Direct is charged only the locations executed before an early abort.
\end{enumerate}
We therefore re-decoded the comparator.

\emph{Decoder.} We use complementary-gap minimum-weight matching~\cite{Bombin2024,GidneyShuttyJones2024} (PyMatching 2.4~\cite{HiggottGidney2025}). The logical observable is appended as an extra detector. Every fault mechanism that fires a postselected cultivation detector is removed from the matching graph, because such a mechanism never occurs alone in a kept shot. The remaining hyperedges are decomposed into existing graphlike mechanisms. For each kept shot, the syndrome is decoded twice, with the observable detector set to 0 and to 1. The prediction is the lighter solution, and the gap is the weight difference in decibels. The postselected set consists of the cultivation detectors other than the eight with coordinate tag $(0,1)$. With this set, the hard-discard rate matches the public data (Table~\ref{tab:3t_validation}).

\begin{table}[H]
\caption{Validation of the re-implemented decoder against the public Sinter data on the same circuit, with idle noise. Our runs use $10^7$ shots at $5\times10^{-4}$ and $4\times10^6$ at $10^{-3}$.}
\label{tab:3t_validation}
\begin{ruledtabular}
\begin{tabular}{l r r r r}
 & \multicolumn{2}{c}{$p=5\times10^{-4}$} & \multicolumn{2}{c}{$p=10^{-3}$} \\
Quantity & this work & public & this work & public \\
\hline
Hard-discard rate & 18.5\% & 18.1\% & 33.6\% & 33.1\% \\
Logical error rate, gap $\ge0$ & $9.6\times10^{-4}$ & $8.1\times10^{-4}$ & $4.1\times10^{-3}$ & $3.6\times10^{-3}$ \\
Logical error rate, gap $\ge25$ & $8.0\times10^{-5}$ & $9.8\times10^{-5}$ & $2.0\times10^{-4}$ & $1.9\times10^{-4}$ \\
Kept fraction at gap $\ge25$ & 0.978 & 0.989 & 0.926 & 0.953 \\
\end{tabular}
\end{ruledtabular}
\end{table}

Our decoder is slightly weaker than the reference decoder, which is conservative for the comparator. To keep the comparison internally consistent, every three-$T$ value below uses our decoder, including the values in the public noise model.

\emph{Early abort.} A rejected attempt is charged the native locations executed up to the tick of its first firing postselected detector, with zero feedback latency. Table~\ref{tab:3t_abort} gives the resulting mean cost per attempt as a fraction $\rho$ of the full path. 

\begin{table}[H]
\caption{Hard-discard rate and early-abort cost fraction $\rho$ of one three-$T$ path, from $4\times10^5$ Stim samples at each point.}
\label{tab:3t_abort}
\begin{ruledtabular}
\begin{tabular}{c r r r r}
 & \multicolumn{2}{c}{with idle noise} & \multicolumn{2}{c}{without idle noise} \\
$p$ & discard & $\rho$ & discard & $\rho$ \\
\hline
$5\times10^{-4}$ & 18.4\% & 0.835 & 9.1\% & 0.918 \\
$8\times10^{-4}$ & 27.9\% & 0.750 & 14.2\% & 0.873 \\
$9\times10^{-4}$ & 30.6\% & 0.726 & 15.8\% & 0.859 \\
$1\times10^{-3}$ & 33.5\% & 0.700 & 17.6\% & 0.843 \\
\end{tabular}
\end{ruledtabular}
\end{table}

\subsection{Output-boundary conventions}

We evaluate the comparator under three output conventions:
\begin{description}
\item[S (public):] the gap is computed from all detectors, including the final noiseless round, and any logical error counts as a failure.
\item[D (Direct):] the final noiseless round is withheld from the gap and serves only for validation. A failure is a closed-boundary logical failure, as in $A_3^{\rm bin}$. A residual detected by the withheld round is an accepted output, not a failure.
\item[P (physical):] the gap is computed from operational detectors only, and a failure is a logical error of a decoder that also sees the continuation.
\end{description}

We also tested adding one noisy distance-13 round (960 locations) before the withheld round, mirroring Direct's faultable expansion round. This variant was never cheaper, so the tables report the variant without the extra round.

Convention D rewards withholding information. With idle noise at $p=10^{-3}$, the closed-boundary logical error rate at gap $\ge0$ is $1.8\times10^{-3}$ without the extra round, with 65\% detectable residuals. With the round it is $3.6\times10^{-3}$, with 45\% detectable residuals. The same property applies to the Direct $A_3^{\rm bin}$.

\subsection{Matched comparison}

At each $p$ and Direct threshold $g$, the comparator uses the cheapest gap setting $g_{3T}$ that meets $3L_T(g_{3T})\le A_3^{\rm bin}(g)\,p^3$. Its cost per $|CS\rangle$ is
\begin{equation}
C_{3T}=3\,\bigl[\rho\cdot1{,}400\bigr]\,N_{g_{3T}}+676,
\end{equation}
where $N_{g}$ is the number of shots per kept $T$ state at gap $\ge g$ (convention D, no idle noise). The comparator runs used $10^7$ shots at $5\times10^{-4}$ and $4\times10^6$ elsewhere, with at least 500 logical errors at every selected operating point. Table~\ref{tab:matched_ler} lists the results; Table~\ref{tab:matched_conventions} shows that the choice of output convention changes the comparator by at most 3\%.

\begin{table}[H]
\caption{Matched comparison, Direct CS with the distance-3 interface. LER target in units of $10^{-4}$. ``Sync.'' is the ratio including synchronous three-factory storage; ``break-even'' is the storage, in $d=13$ patch-rounds per $|CS\rangle$, above which Direct CS is cheaper (0 when it is cheaper without storage).}
\label{tab:matched_ler}
\begin{ruledtabular}
\small
\begin{tabular}{c c c c r r c c c}
Threshold & $p$ & Target & $g_{3T}$ & $C_{\rm Direct}$ & $C_{3T}$ & Ratio & Sync. & Break-even \\
\hline
$g=10$ & $5\times10^{-4}$ & $1.15$ & 21 & 5,227 & 5,167 & 1.01 & 0.49 & 0.06 \\
$g=10$ & $8\times10^{-4}$ & $4.70$ & 7 & 6,451 & 5,211 & 1.24 & 0.53 & 1.29 \\
$g=10$ & $9\times10^{-4}$ & $6.69$ & 5 & 6,824 & 5,153 & 1.32 & 0.59 & 1.74 \\
$g=10$ & $1\times10^{-3}$ & $9.18$ & 5 & 7,344 & 5,180 & 1.42 & 0.60 & 2.25 \\
$g^\ast=4$ & $5\times10^{-4}$ & $2.13$ & 20 & 5,056 & 5,162 & 0.98 & 0.47 & 0.00 \\
$g^\ast=4$ & $8\times10^{-4}$ & $8.71$ & 4 & 5,904 & 5,075 & 1.16 & 0.57 & 0.86 \\
$g^\ast=4$ & $9\times10^{-4}$ & $12.40$ & 0 & 5,953 & 4,965 & 1.20 & 0.66 & 1.03 \\
$g^\ast=4$ & $1\times10^{-3}$ & $17.01$ & 0 & 6,341 & 4,969 & 1.28 & 0.67 & 1.43 \\
\end{tabular}
\end{ruledtabular}
\end{table}

\begin{table}[H]
\caption{Comparator cost under the three output conventions, at the original $A_3^{\rm bin}p^3$ target of the distance-5 interface, with and without idle noise. Costs are active locations per $|CS\rangle$.}
\label{tab:matched_conventions}
\begin{ruledtabular}
\small
\begin{tabular}{c l c r r r c}
$p$ & 3T noise & Conv. & Gap & $L_T$ & $N_g$ & $C_{3T}$ \\
\hline
$5\times10^{-4}$ & with idle & S & 25 & $7.98\times10^{-5}$ & 1.255 & 5,079  \\
$5\times10^{-4}$ & with idle & D & 23 & $8.00\times10^{-5}$ & 1.426 & 5,677  \\
$5\times10^{-4}$ & with idle & P & 25 & $8.03\times10^{-5}$ & 1.460 & 5,797  \\
$5\times10^{-4}$ & no idle & S & 0 & $7.53\times10^{-5}$ & 1.101 & 4,924  \\
$5\times10^{-4}$ & no idle & D & 5 & $9.22\times10^{-5}$ & 1.128 & 5,027  \\
$5\times10^{-4}$ & no idle & P & 0 & $7.05\times10^{-5}$ & 1.101 & 4,924  \\
$8\times10^{-4}$ & no idle & S & 0 & $1.98\times10^{-4}$ & 1.167 & 4,955  \\
$8\times10^{-4}$ & no idle & D & 0 & $3.64\times10^{-4}$ & 1.167 & 4,953  \\
$8\times10^{-4}$ & no idle & P & 0 & $1.92\times10^{-4}$ & 1.167 & 4,953  \\
$9\times10^{-4}$ & no idle & S & 0 & $2.65\times10^{-4}$ & 1.189 & 4,964  \\
$9\times10^{-4}$ & no idle & D & 0 & $4.07\times10^{-4}$ & 1.189 & 4,965  \\
$9\times10^{-4}$ & no idle & P & 0 & $2.63\times10^{-4}$ & 1.189 & 4,965  \\
$1\times10^{-3}$ & with idle & S & 19 & $6.83\times10^{-4}$ & 1.557 & 5,253  \\
$1\times10^{-3}$ & with idle & D & 8 & $6.77\times10^{-4}$ & 1.782 & 5,918  \\
$1\times10^{-3}$ & with idle & P & 20 & $6.10\times10^{-4}$ & 2.056 & 6,721  \\
$1\times10^{-3}$ & no idle & S & 0 & $3.26\times10^{-4}$ & 1.213 & 4,968  \\
$1\times10^{-3}$ & no idle & D & 0 & $4.64\times10^{-4}$ & 1.213 & 4,969  \\
$1\times10^{-3}$ & no idle & P & 0 & $3.30\times10^{-4}$ & 1.213 & 4,969  \\
\end{tabular}
\end{ruledtabular}
\end{table}

\subsection{Decoder-threshold selection}

The selection rule and the calibration data were fixed before any calibration outcome was seen. The calibration data are finite-$p$ runs at $5\times10^{-4}$ and $10^{-3}$ (5,000 shots each, prior seed 30270901, evaluation seed 30270921) and 101,864 order-three realizations (seed 30281021). These seeds are disjoint from every reported result. Table~\ref{tab:gap_selection} gives the calibration values. The rule selects $g^\ast=4$. Reported values at $g^\ast$ come from the independent reporting data.

\begin{table}[H]
\caption{Threshold selection on calibration data. $A_3^{\rm bin}$ in units of $10^6$; ratios are $C_{\rm Direct}/C_{3T}(A_3^{\rm bin}p^3)$; $\bar R$ is their mean.}
\label{tab:gap_selection}
\begin{ruledtabular}
\begin{tabular}{c c c c c}
Gap & $A_3^{\rm bin}$ & Ratio, $5\times10^{-4}$ & Ratio, $10^{-3}$ & $\bar R$ \\
\hline
4 & $1.82$ & 0.983 & 1.262 & 1.122 \\
5 & $1.73$ & 0.981 & 1.269 & 1.125 \\
6 & $1.59$ & 0.980 & 1.300 & 1.140 \\
7 & $1.59$ & 0.984 & 1.393 & 1.189 \\
8 & $1.32$ & 1.008 & 1.387 & 1.197 \\
9 & $1.32$ & 1.010 & 1.415 & 1.212 \\
10 & $1.32$ & 1.014 & 1.421 & 1.218 \\
12 & $1.32$ & 1.016 & 1.429 & 1.223 \\
\end{tabular}
\end{ruledtabular}
\end{table}

\subsection{Zero-noise decomposition}

Table~\ref{tab:p0_decomposition} compares the two routes at $p=0$, where every attempt is accepted.

\begin{table}[H]
\caption{Active locations per $|CS\rangle$ at $p=0$, native convention.}
\label{tab:p0_decomposition}
\begin{ruledtabular}
\small
\begin{tabular}{l r r r}
Stage & Direct & Three-$T$ & Difference \\
\hline
Source core & 890 & $3\times236=708$ & $+182$ \\
Steane$\rightarrow d=3$ escape & $2\times316=632$ & -- & $+632$ \\
Growth to $d=13$ & $2\times1{,}120=2{,}240$ & $3\times1{,}164=3{,}492$ & $-1{,}252$ \\
Conversion to $|CS\rangle$ & -- & 676 & $-676$ \\
\hline
Total & 3,762 & 4,876 & $-1{,}114$ \\
\end{tabular}
\end{ruledtabular}
\end{table}

With the distance-5 interface, the escape and two distance-5 rounds cost $2\times1{,}028$ and the Direct total was 5,154.

\subsection{$T$-state storage}

Direct CS produces both output patches in a single accepted attempt. The three-$T$ route must hold finished $T$ states as distance-13 patches until all three are available. One distance-13 round costs 960 native locations (624 CX, 168 R, and 168 M) and lasts 8 circuit ticks; a full $T$ attempt lasts 89 ticks. Let $\bar t_T$ be the mean time per kept $T$ state from one factory, including aborted attempts. We consider two schedules:
\begin{itemize}
\item \emph{Synchronous assembly:} three factories run in parallel and stop on success. The holding cost is $\sum_i(\max_j t_j-t_i)$, evaluated by Monte Carlo from the empirical discard rates, mean abort time, and $N_g$.
\item \emph{Buffered pipeline:} $k$ factories feed one assembler. With Poisson arrivals at rate $k/\bar t_T$, the buffer passes through occupancies 1 and 2 before each assembly, so the mean holding time per $|CS\rangle$ is $3\bar t_T/k$.
\end{itemize}
At $g^\ast$, the break-even factory counts are about 42, 35, and 25 at $p=8\times10^{-4}$, $9\times10^{-4}$, and $10^{-3}$. At $5\times10^{-4}$ Direct CS is cheaper at any $k$. The holding rounds add negligible logical error at distance 13; they are charged as cost only.

\subsection{Spacetime}

We count qubit-ticks per accepted $|CS\rangle$.
\begin{itemize}
\item \emph{Comparator.} Each $T$ attempt is charged the sum over qubits of the ticks between the qubit's first and last operation, truncated at the first firing postselected detector: 4,664 per full attempt, and 4,301--3,967 per attempt on average over $5\times10^{-4}$--$10^{-3}$. Conversion is charged as three $d=13$ patches for four ticks.
\item \emph{Direct downstream.} Charged by the same lifetime rule from its Stim circuit: 2,950 per patch.
\item \emph{Direct front end and core.} These are not Stim circuits. They use their native schedule (356 and 2,040), which bounds the lifetime count from above. Front-end and core reach fractions per attempt come from 10,000 fresh samples at each $p$.
\end{itemize}

In the convention least favorable to Direct CS, the downstream is charged as a rectangle of touched qubits over all ticks: $33\times32+2{,}696=3{,}752$ per patch. The comparator is then charged its published successful-path volume of 3,690 per $T$ state, with rejected attempts free, and the conversion counts data qubits only. Table~\ref{tab:spacetime} gives both.

\begin{table}[H]
\caption{Qubit-ticks per accepted $|CS\rangle$, without storage. ``Same rule'': lifetime counting for both Stim circuits. ``Least favorable'': rectangle counting for the Direct downstream and published successful-path volume for the comparator.}
\label{tab:spacetime}
\begin{ruledtabular}
\small
\begin{tabular}{c c r r c r r c}
 & & \multicolumn{3}{c}{Same rule} & \multicolumn{3}{c}{Least favorable to Direct} \\
Threshold & $p$ & Direct & 3T & Ratio & Direct & 3T & Ratio \\
\hline
$g=10$ & $5\times10^{-4}$ & 11,943 & 19,066 & 0.63 & 14,030 & 13,901 & 1.01 \\
$g=10$ & $8\times10^{-4}$ & 14,392 & 19,262 & 0.75 & 16,729 & 13,934 & 1.20 \\
$g=10$ & $9\times10^{-4}$ & 15,250 & 19,055 & 0.80 & 17,674 & 13,753 & 1.29 \\
$g=10$ & $1\times10^{-3}$ & 16,944 & 19,190 & 0.88 & 19,565 & 13,812 & 1.42 \\
$g^\ast=4$ & $5\times10^{-4}$ & 11,553 & 19,050 & 0.61 & 13,572 & 13,888 & 0.98 \\
$g^\ast=4$ & $8\times10^{-4}$ & 13,172 & 18,806 & 0.70 & 15,310 & 13,582 & 1.13 \\
$g^\ast=4$ & $9\times10^{-4}$ & 13,303 & 18,423 & 0.72 & 15,418 & 13,267 & 1.16 \\
$g^\ast=4$ & $1\times10^{-3}$ & 14,630 & 18,479 & 0.79 & 16,894 & 13,267 & 1.27 \\
\end{tabular}
\end{ruledtabular}
\end{table}

\subsection{Scope}

The comparison in this Note uses active locations and qubit-ticks without idle noise for either route; Supplementary Note~10 adds idle noise. A buffered pipeline lowers the comparator's storage by adding factories, whose footprint neither metric charges. Three assumptions favor the comparator: routing between factories and assembler is free, the conversion is charged as transversal gates without intervening rounds, and its rejected attempts are charged up to the first firing postselected detector, whereas Direct attempts are charged per stage. Three choices favor Direct: the closed-boundary failure definition, the single-location CZZ convention, and the simulation of the comparator with a separate noise location per Hadamard. Because the comparator needs little gap postselection, dropping the quality target entirely lowers its cost by at most 2\%. The three-$T$ logical error rates come from a Y-basis surrogate circuit~\cite{Gidney2024Ybasis}, as in Ref.~\cite{Sahay2026}.

\section{Supplementary Note 9: Downstream Schedule Anatomy}
\label{supp:metric_sensitivity}

\begin{figure}[H]
\centering
\includegraphics[width=0.95\textwidth]{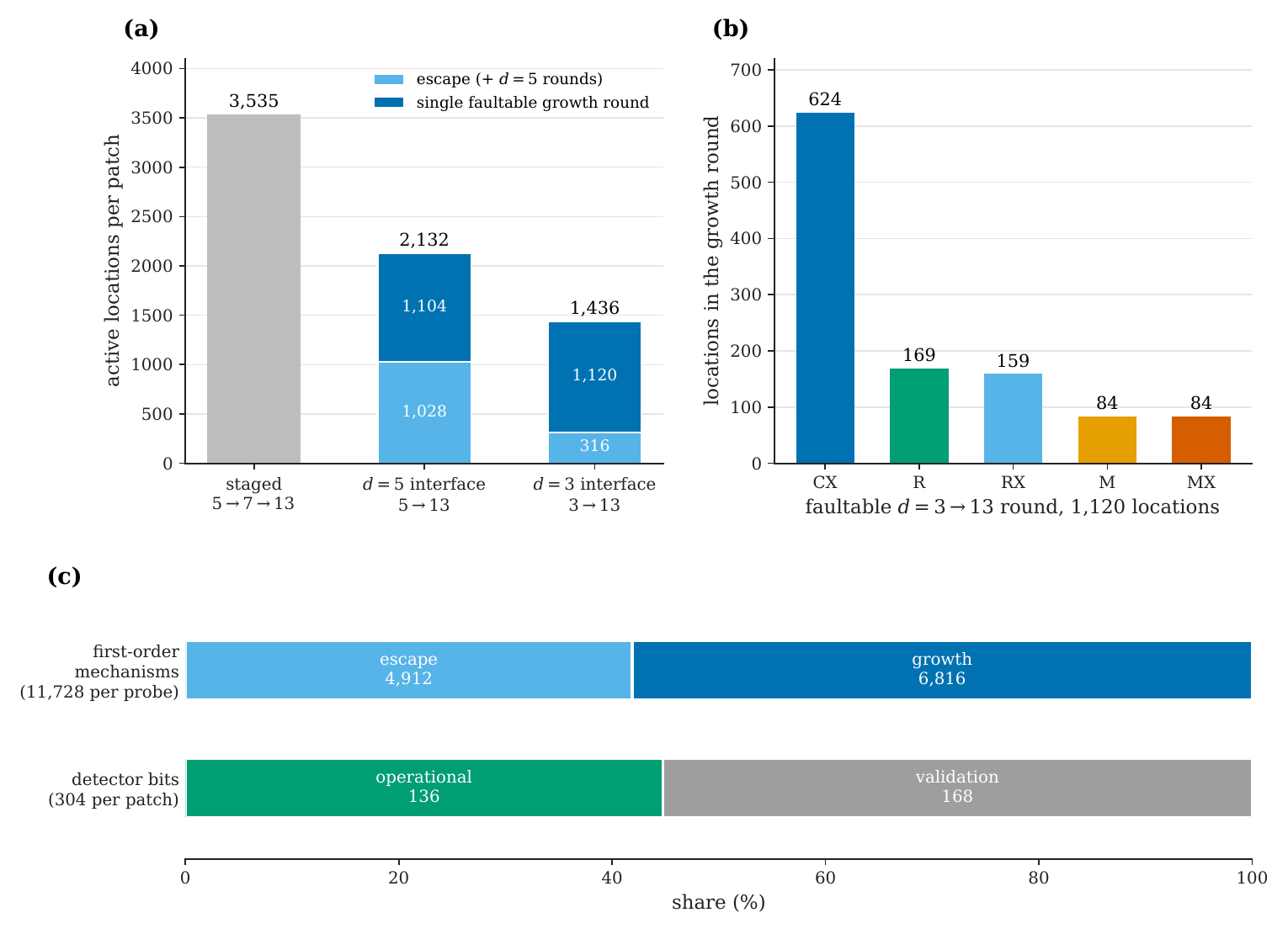}
\caption{\textbf{Downstream schedule.} (a) Active locations per patch for the earlier staged schedule (3,535), the distance-5 single-round interface (2,132), and the distance-3 interface (1,436). (b) Gate inventory of the faultable $d=3\rightarrow13$ expansion round. (c) First-order mechanism counts for one recovery probe and the 136/168 operational/validation detector split. Mechanism counts are not probability weighted.}
\label{fig:supp_expansion_anatomy}
\end{figure}

Per output patch, the Steane-to-$d=3$ escape contributes 316 active locations and the faultable direct $d=3\rightarrow13$ round contributes 1,120. The expansion-round inventory is 624 CX, 169 R, 159 RX, 84 M, and 84 MX. For each individual X- or Z-recovery probe, the structural catalogue contains 11,728 concrete first-order Pauli mechanisms: 4,912 from the escape and 6,816 from the expansion. These are mechanism counts, not probability weights. After joint X/Z merging, the complete set contains 12,734 first-order mechanisms per patch. The detector interface contains 136 operational and 168 future-validation bits.

\section{Supplementary Note 10: Idle-Memory Noise}

\subsection{Idle rule and schedule}
We add idle-memory noise to both routes with a single rule. A qubit is \emph{alive} from its preparation to its last operation, excluding the interval between a measurement and a subsequent reset, which carries no state. In every layer in which an alive qubit is not acted on, it receives single-qubit depolarizing noise of strength $p$. The rule is stricter than the idle model of the public comparator circuit of Ref.~\cite{Sahay2026}, which leaves some waiting qubits noiseless during the cultivation stage. Under our rule the public comparator schedule has 1,951 idle locations per $T$ path, and its hard-discard rate at $p=5\times10^{-4}$ is 27.5\% instead of 18.4\%.

Idle counts depend on the schedule, and the Direct front end and core had no layer schedule. We therefore schedule both routes by the same procedure. Each qubit keeps its order of operations, and only the layer index $t_i$ of each operation is chosen. We minimize the total number of idle locations,
\begin{equation}
\min_{t}\ \sum_{q}\sum_{\text{segments of }q}\bigl(t_{\rm last}-t_{\rm first}\bigr)
\quad\text{subject to}\quad t_j\ge t_i+1,
\end{equation}
for every pair of consecutive operations $i\to j$ on a common qubit.
All constraints are differences of two variables, so the constraint matrix is totally unimodular and the linear-programming optimum is integral. The circuit is unchanged by construction: every qubit sees the same sequence of operations, so all certificates of the main text apply verbatim. For the Direct front end and core we further require identical internal timing for every instance of a gadget (encoder, error-detection block, CAT preparation, logical-check body per pair). This raises the optimum from 204 to 207 idle locations.

\begin{table}[h]
\caption{Idle locations under the common rule.}
\begin{ruledtabular}
\small
\begin{tabular}{l r r r}
 & Active & Idle, original & Idle, retimed \\
\hline
Direct front end + core & 890 & $\approx$1,110$^a$ & 207 \\
Direct downstream, per patch & 1,436 & 283 & 233 \\
Three-$T$ comparator, per $T$ path & 1,400 & 1,951 & 1,000 \\
\end{tabular}
\end{ruledtabular}
\footnotesize $^a$Native rectangle schedule.
\end{table}

\subsection{Simulation}
The idle locations are simulated explicitly. In the Direct front end and core they are inserted as additional single-qubit Pauli locations into the state-vector Monte Carlo: in the encoder and CAT preparation as propagated Pauli frames, in error detection as identity operations, and in the logical-check bodies as identity gates. Where the retimed schedule differs between instances of a shared gadget, the maximum is used, giving 266 modelled idle locations for 207 scheduled ones. In the downstream circuit they are \texttt{DEPOLARIZE1} instructions at the retimed positions. The error-detection single-fault self-test passes with the idle operations included (812 single faults, exact and propagated actions agree). Including the downstream idle locations leaves the set of joint X/Z effect classes unchanged at 1,071: no idle fault produces a symptom that an existing single location does not. Consequently the operational firewall and the low-order audit are unchanged, and the sampled closed-boundary harmful count at total fault order $\le2$ is zero at both error rates. The decoder and its thresholds are unchanged. The resource ledger charges active locations only, as without idle noise.

\subsection{Result}
Idle noise lowers the Direct acceptance from 0.545 to 0.441 at $p=5\times10^{-4}$ and from 0.335 to 0.229 at $10^{-3}$ ($g^\ast=4$), raising the cost per accepted output by 15\% and 24\%. The comparator, re-decoded with the same complementary-gap decoder on its retimed circuit, rises by 4\% and 8\%. Its cost is dominated by locations, not acceptance. The order-three coefficient with idle noise was not re-estimated. The comparator target uses $A_3^{\rm bin}$ multiplied by $k=1.6$, the growth of the number of location triples in the dominant downstream strata, and Table~\ref{tab:idle_compare} shows the sensitivity to $k$.

\begin{table}[h]
\caption{Active locations per accepted output with idle noise ($g^\ast=4$). $C_{\rm 3T}$ at $k=1.6$; ``Storage'' is synchronous three-factory assembly ($k=1.6$); ``No idle'' is the main-text ratio without idle noise. Ratios below one favor Direct CS.}
\label{tab:idle_compare}
\begin{ruledtabular}
\small
\begin{tabular}{c r r r r r}
$p$ & $C_{\rm Direct}$ & $C_{\rm 3T}$ & Ratio ($k=1,\,1.6,\,2.5$) & Storage & No idle\\
\hline
$5\times10^{-4}$ & $5{,}796\pm92$ & 5,391 & 1.04, 1.08, 1.09 & 0.37 & 0.98\\
$10^{-3}$ & $7{,}874\pm205$ & 5,385 & 1.36, 1.46, 1.54 & 0.42 & 1.28\\
\end{tabular}
\end{ruledtabular}
\end{table}

Idle noise therefore widens the active-location gap in favor of the comparator by about 10 points at $5\times10^{-4}$ and 18 points at $10^{-3}$. Direct CS is cheaper whenever the comparator holds more than 0.4 and 2.6 $d=13$ patch-rounds per $|CS\rangle$ at the two error rates, and it remains cheaper under synchronous three-factory assembly. The same retiming also reduces the comparator's idle load by half, and against the original public schedule the ratios are 1.02 and 1.29.

\paragraph*{Limitations.} All operations, including measurement and reset, occupy one layer. On hardware where measurement and reset are several times slower than gates, the idle load grows for both routes, and more for Direct CS, whose core contains more mid-circuit measurements (CAT verification and error-detection ancillas). Duration-weighted constraints $t_j\ge t_i+\tau_i$ keep the retiming problem a difference-constraint linear program. That case is left for future work.

\bibliographystyle{unsrtnat}
\bibliography{direct_cs_references}